\documentclass[a4paper,11pt]{article}
\usepackage{jheppub} 

\usepackage{microtype} 

\usepackage{comment}
\usepackage{dcolumn}
\usepackage{xcolor}
\usepackage{youngtab}
\usepackage{ytableau}
\usepackage{tabularray}
\usepackage{booktabs}    
\usepackage{placeins}
\usepackage{soul}
\usepackage{bm}
\definecolor{redd}{rgb}{0.8, 0.1,0.2}
\definecolor{navy}{rgb}{0.05, 0.23,0.75}
\usepackage{chngcntr}
\hypersetup{colorlinks,	linkcolor={green!47!black},	citecolor={green!60!black},	urlcolor={green!55!black}}
\usepackage{bbm}
\usepackage{amsfonts}
\usepackage{amsmath,amssymb}
\usepackage{mathrsfs}
\usepackage{epsfig}
\usepackage{graphicx}
\usepackage{wrapfig}                 
\usepackage{url}
\usepackage[nameinlink]{cleveref}
\usepackage{float}
\usepackage{color}
\usepackage{multirow}
\usepackage{lipsum}
\usepackage{enumitem}
\newcolumntype{L}{>{\centering\arraybackslash}m{1.5cm}}

\usepackage{enumitem}
\usepackage{chngcntr}
\usepackage{tikz}
\usepackage{adjustbox}
\usepackage{cancel}

\newcommand{\be}{\begin{equation}}
\newcommand{\ee}{\end{equation}}
\newcommand{\bea}{\begin{eqnarray}}
\newcommand{\eea}{\end{eqnarray}}
\newcommand{\bc}{\begin{center}}
\newcommand{\ec}{\end{center}}

\newcommand{\Ngen}{N_{\textrm{gen}}}

\newcommand{\kSSB}{k_{\textrm{dSB}}}

\newcommand{\dSSB}{\textrm{d}\textrm{SB}}
\newcommand{\acrit}{\alpha^\textrm{crit}_{\textrm{SB}}}

\newcommand{\orcid}[1]{\href{https://orcid.org/#1}{	\raisebox{0.5\height}{\includegraphics[height=1.25ex,width=1.25ex]{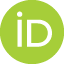}}}}

\graphicspath{{Figures/}}

\makeatletter \gdef\@fpheader{} \makeatother

\begin{document}
		
\preprint{RIKEN-iTHEMS-Report-25}

\title{Dynamical symmetry breaking in Georgi-Glashow chiral-gauge theories}

\author{Hao-Lin Li\,\orcid{0000-0002-4310-1915}}
\affiliation{School of Physics, Sun Yat-Sen University, Guangzhou 510275, P. R. China}

\author{\' Alvaro Pastor-Guti\'errez\,\orcid{0000-0001-5152-3678}}
\affiliation{RIKEN Center for Interdisciplinary Theoretical and Mathematical Sciences (iTHEMS), RIKEN, Wako 351-0198, Japan}

\author{Shahram Vatani\,\orcid{0000-0002-1277-5829}}
\affiliation{ Centre for Cosmology, Particle Physics and Phenomenology (CP3), Université catholique de Louvain, Chemin du Cyclotron, 2 B-1348 Louvain-la-Neuve, Belgium}

\author{Ling-Xiao Xu\,\orcid{0000-0002-4970-2404}\,}
\affiliation{Abdus Salam International Centre for Theoretical Physics, Strada Costiera 11, 34151, Trieste, Italy}

\abstract{
We investigate dynamical symmetry breaking in a class of chiral gauge theories containing the Georgi-Glashow model. These theories feature a gauge sector and two fermion species that transform in the two-index antisymmetric and antifundamental representations with different multiplicities. Using the effective action formalism and the functional renormalization group, we derive the flow of four-fermion interactions that encode their resonant structure and information about bound-state formation. Generalizing the theories to multiple generations, we make contact with the loss of asymptotic freedom and dissect the boundary of a conjectured conformal window. 
Our results show that, while most of the theory space displays a dominant color-breaking condensate, there exists a strongly coupled regime where the lowest-laying mechanisms fail and more intricate dynamics are expected to arise. This analysis provides a first step toward the infrared behavior of chiral gauge theories with functional methods.
}

\maketitle

\section{Introduction}\label{sec:Introduction}
One of the outstanding challenges in Quantum Field Theory (QFT) and particle physics is to systematically understand the infrared (IR) phases of gauge theories in $(3+1)$ spacetime dimensions. A wide array of approaches, both exact and approximate, have been developed to probe the phases of such theories.
Constraints have been developed from tools as the ’t Hooft anomaly matching conditions~\cite{tHooft:1979rat} or via connections to the supersymmetric limit~\cite{Seiberg:1994rs,Seiberg:1994aj,Seiberg:1994bz,Seiberg:1994pq}. For vector-like gauge theories, the fermion path integral measure can be made real and positive by appropriately choosing fermion mass terms, thereby permitting the application of additional criteria such as the persistent mass condition\footnote{See e.g.~\cite{Ciambriello:2022wmh, Ciambriello:2024xzd} for recent studies on the use of persistent mass conditions together with 't Hooft anomaly matching conditions in a class of QCD-like theories, where it is proven that chiral symmetry must be spontaneously broken once the theory flows to an IR-free fixed point with full screening of color charges.}~\cite{Vafa:1983tf}. Moreover, major insights into physical correlations in the strongly coupled regime have been obtained from first-principle computations such as Monte Carlo lattice simulations (see e.g.~\cite{Wilson:1974sk, Kogut:1982ds, Greensite:2003bk, Bogolubsky:2009dc, Bowman:2004jm, Aoki:2009sc}) and functional methods (see e.g.~\cite{Alkofer:2000wg, Fischer:2006ub, Pawlowski:2003hq, Braun:2007bx, Cyrol:2016tym, Aguilar:2022thg}).

Although most efforts have focused on gauge theories with vector-like fermions, due to its relevance to Quantum Chromodynamics (QCD)~\cite{Gross:2022hyw}, the study of chiral gauge theories remains of significant theoretical and phenomenological interest. On the one hand, a deeper understanding of chiral gauge theories is essential, as the Standard Model (SM) of particle physics is constructed based on a chiral matter content. However, strongly coupled chiral gauge theories~\cite{Raby:1979my, Dimopoulos:1980hn, Bars:1981se} may potentially be relevant for addressing outstanding questions beyond the SM. However, our limited understanding of their strongly coupled behavior represents a major obstacle to assessing their phenomenological viability. Accordingly, it is of paramount importance to advance our knowledge of the dynamics governing these theories.

By construction, no gauge-invariant fermion bilinear mass terms exist in chiral-gauge theories, which renders the persistent mass conditions inapplicable. Furthermore, the lattice regularization of chiral fermions remains a subtle and technically challenging task due to the Nielsen–Ninomiya theorem~\cite{Nielsen:1980rz, Nielsen:1981hk, Nielsen:1981xu}. 
These challenges have motivated a long-standing effort to develop robust theoretical tools capable of systematically probing the IR phases of chiral gauge theories. Recent progress has been made by exploiting ’t Hooft anomaly matching conditions associated with generalized global symmetries; for reviews, see e.g.~\cite{Bolognesi:2021jzs, Bolognesi:2023sxe, Konishi:2024rjz}. Despite these advances, our understanding of chiral gauge dynamics remains incomplete, as either analytical or numerical methods are still largely unavailable.

The IR phases of such theories are expected to depend crucially on the fermion content \cite{Sannino:2009za,Poppitz:2009uq,Bolognesi:2019wfq,Bolognesi:2020mpe,Bolognesi:2021yni,Bolognesi:2021jzs,Bolognesi:2023sxe,Bolognesi:2024bnm}. Some models may flow to an interacting IR fixed point, while others are conjectured to exhibit non-trivial Higgsing dynamics with condensate formation, leading to rich phenomena such as color-flavor locking, tumbling, and dynamical Abelianization. The standard lore is that one can smoothly interpolate between the Higgs and confinement regimes without encountering a phase transition~\cite{Fradkin:1978dv,Dimopoulos:1980hn}.

In this paper, we study a class of anomaly-free and purely chiral gauge theories of the Georgi-Glashow (GG) type using the functional renormalization group (fRG) approach~\cite{Wetterich:1992yh,Dupuis:2020fhh}. Due to its non-perturbative nature, the fRG is particularly suitable for investigating the strongly coupled dynamics of these theories. In this first work, we focus on the dynamical symmetry breaking ($\dSSB$) mechanisms which give rise to the highest condensate in the theory following in spirit previous studies in QCD-like theories~\cite{Miransky:1998dh, Miransky1997, Gies:2005as, Braun:2006jd, Braun:2009ns, Goertz:2024dnz, Braun:2010qs}. While here we focus on the start of dynamics, the present approach has proven its quantitative precision to describe dynamical chiral symmetry breaking and color confinement. Remarkably, it is shown in~\cite{Cyrol:2017ewj,Mitter:2014wpa,Ihssen:2024miv,Goertz:2024dnz,Fu:2019hdw} that the fRG approach can achieve quantitative precision with lattice results where they exist. Consequently, no conceptual complication is expected in the present approach.

In this work, we derive the RG flows for the dressings of the four-fermion operators which provide us with knowledge on the channel's resonant structure and the critical coupling necessary to trigger a dynamical scale generation. To the best of our knowledge, this class of multispecies gauge–fermion theories, where each fermion transforms under distinct, nonconjugate representations of the gauge group, has not been previously studied using functional methods. This provides us with unprecedented information on the properties of the theory at the start of non-perturbative dynamics as well as allows us to chart theory space where various regions with distinct dynamical patterns can be identified, including a conjectured conformal limit. 
 
The remainder of the paper is organized as follows. In~\Cref{sec:GG} we introduce the Georgi-Glashow model and its generalized form, which allows us to make contact to the loss of asymptotic freedom and a quantum scale-invariant regime. In~\Cref{sec:dSSBandEA} we introduce the effective action approach, the methodology for diagnosing $\dSSB$, and the fRG approach. In~\Cref{sec:phasediagram} we discuss the results of the flow equation, concerning the resonant structure of the different channels and the phases in theory space, including the quantum scale invariant limit. Before concluding, we speculate about the IR fate of the theories and discuss the connection to other studies. Technical details concerning the basis construction and the flow equations as well as single examples can be found in the Appendices.

\section{The chiral theories}\label{sec:GG}
We are interested in purely chiral theories and in this work we focus on a general class consisting of $SU(N_c)$ gauge theories  ($N_c\geq 5$) with left-handed Weyl fermions in the respective complex representations,
\begin{align}
\Ngen \; {\tiny\yng(1,1)} \;\; \bm{\oplus}\;\; \Ngen \left(N_c - 4 \right) \; \overline{{\tiny\yng(1)}} \;,
\end{align}
where the fermion multiplicities are related in order to cancel the $[SU(N_c)]^3$ triangle gauge anomaly. When $N_c=5$, the fermion content coincides with the model originally proposed by Georgi and Glashow in the context of grand unification of the SM gauge couplings~\cite{Georgi:1974sy}. Due to this, models of this type with $\Ngen=1$ are later referred to as chiral-gauge theories of the Georgi-Glashow type~\cite{Raby:1979my, Dimopoulos:1980hn}, arguably the simplest purely chiral models. 

Motivated by the structural similarity to the SM, throughout the rest of the paper we refer to $\Ngen$ as the number of \emph{generations} or copies of the minimal GG theory. Within each generation, the model contains one Weyl fermion in the two-index antisymmetric representation, $\chi$, and accordingly, $(N_c-4)$ Weyl fermions in the anti-fundamental representation, $\psi$. The GG model with $\Ngen$ generations of chiral fermions has the global flavor symmetry group 
\begin{align}
G_f=SU\left(\Ngen\left( N_c - 4\right)\right)\times SU\left(\Ngen\right) \times U\left(1\right)\; ,
\label{eq:flavor_sym}
\end{align}
up to some central group elements acting trivially on the chiral fermions; see \Cref{table:GGcontent} for the various fermion charges. Notice that $G_f$ reduces to a single $U(1)$ when $N_c=5$ and $\Ngen=1$, otherwise there are always non-Abelian factors. 

\begin{table}[t!]
\centering
\renewcommand{\arraystretch}{2}
\begin{tabular}{|l||c||c|c|c|} 
\hline
{} & $SU(N_c)$ & $\, \,SU\left(\Ngen (N_c-4)\right)\,\,$  & $\,\,SU\left(\Ngen\right)\,\,$ &  $U(1)$   \\ 
\hline 
$\psi$ & $\overline{{\tiny\yng(1)}}$ & \tiny{$\yng(1)$}  & {$1$} &{ $-(N_c -2)$ } \\
\hline
$\chi$ & $\tiny\yng(1,1)$ & {$1$} & \tiny{$\yng(1)$}  & {$ N_c-4$} \\
\hline
\end{tabular}
\caption{Fermion content of the generalized GG model with $\Ngen$ generations, where the fermion quantum numbers are specified under the $SU(N_c)$ gauge group and the flavor symmetry group defined in Eq.~\eqref{eq:flavor_sym}. Here, the $U(1)$ factor is free from the $U(1)[SU(N_c)]^2$ Adler-Bell-Jackiw anomaly.}
\label{table:GGcontent}
\end{table}

Theories with $\Ngen < \Ngen^{\rm AF}= (11 N_c)/(2N_c - 6)$ have a well-defined asymptotically free ultraviolet (UV) limit. When $\Ngen$ is slightly below~$\Ngen^{\rm AF}$, it is believed that theories flow into an interacting IR fixed point, analogous in spirit to the Caswell-Banks-Zaks~\cite{Caswell:1974gg, Banks:1981nn} solutions in QCD-like theories. However, unlike the latter, determining the existence of the fixed point in the GG-type models is tricky. In particular, a perturbative analysis of the existence of a fixed point is not valid, since the zeros of the $\beta$ function cannot be \emph{parametrically} under control\footnote{It may still be \emph{numerically} small, but this is not sufficient to prove the existence of an interacting fixed point in the perturbative regime.}, contrary to QCD-like theories in the Veneziano-Witten limit. 
A simple estimation using the leading two coefficients of the $\beta$ function suggests that all the interacting fixed points, if they exist at all, are non-perturbative, even when the gauge coupling can be numerically small; see~\cite{Bolognesi:2023sxe} for a review emphasizing this point. For sufficiently small $\Ngen$, fixed-point solutions are not found and the gauge coupling drives the appearance of so-far unknown dynamics which we target in this work. 

We note that while the strongly coupled regime of the one-generation model was studied in~\cite{Raby:1979my, Dimopoulos:1980hn} long ago, the multi-generation GG models, which we refer to as the \textit{generalized}, have received very little attention; for instance, see only~\cite {Bolognesi:2019wfq, Bai:2021tgl} for some recent studies.

\section{Effective action and the functional approach}\label{sec:dSSBandEA}
In the present work, we employ the effective action formalism which encodes all physical information in the one-particle irreducible generating functional, from which correlation functions can be extracted. These fully-dressed correlators or $n$-point functions capture the relevant dynamics of the theory, including known phenomena like confinement and dynamical chiral symmetry breaking, as observed in QCD, as well as potential new mechanisms not yet identified. A particularly important feature in gauge–fermion quantum field theories is the dynamical formation of a condensate, which leads to the generation of an explicit mass scale. In the present classically scale-invariant theories, this also encodes the breaking of quantum scale invariance, which may otherwise be given in the presence of an IR fixed point, as in the CBZ scenario.

The dynamical emergence of fermion condensates has a long history in the context of effective actions, particularly in condensed matter and nuclear physics~\cite{Nambu:1961fr,Braun:2011pp}. This mechanism is encoded in the fermionic $n$-point functions and can commonly be diagnosed in a divergence of the dressings. 
Through a Hubbard–Stratonovich transformation~\cite{HubbardPhysRevLett.3.77,Stratonovich}, this divergence is shown to be mathematically equivalent to the appearance of a condensate in the bosonized formulation of the theory. In that setting, the formation of a condensate corresponds to a change in the curvature of the bosonic effective potential around the origin, specifically, a change in the sign of the boson mass term. Thus, the transition from a symmetric to a broken phase is indicated by a vanishing boson mass (i.e., an infinite correlation length), or equivalently, by a divergence in the corresponding four-fermion coupling. See, e.g.,~\cite{Miransky1989b,Miransky:1994vk,Gies:2005as} for illustrative examples.

In this work, we follow these lines to analyze the appearance of condensates in generalized GG models. To account for quantum corrections and derive the $n$-point functions of the effective action, we employ the functional renormalization group~\cite{Wetterich:1992yh}. In this non-perturbative approach, momentum modes in the path integral measure are integrated out progressively with the help of an IR regulator function $R_k$, which suppresses fluctuations below the cutoff scale $k$. This procedure defines the effective average action $\Gamma_k$~\cite{Wetterich:1989xg,Wetterich:1991be}, which incorporates quantum fluctuations down to the momentum scale $k$ and recovers the full effective action in the limit $k \to 0$. Central to this approach is the flow or Wetterich equation with which we implement quantum corrections in a non-perturbative manner,
\begin{align}\label{eq:floweq}
	\partial_t \Gamma_k \left[\Phi\right]= \frac{1}{2}\,{\rm STr} \left[ \left(\Gamma_k^{(2)}\left[\Phi\right]+R_k\right)^{-1} \cdot \partial_t R_k \right]\,.
\end{align}
Here, $\Phi$ denotes the superfield containing all fields in the theory, $\partial_t \equiv k\, \partial_k$ is the RG-time derivative, and $\Gamma_k^{(2)}\left[\Phi\right]$ is the second functional derivative of $\Gamma_k\left[\Phi\right]$, i.e. the full inverse propagator Hessian. The supertrace $\mathrm{STr}$ includes a trace over momentum space as well as all internal (gauge and global) indices, with a minus sign for fermionic degrees of freedom.

Exactly solving the flow equation \eqref{eq:floweq} is generally not feasible, and practical applications rely on truncations of the effective average action. These consist of considering a finite set of operators consistent with the symmetries of the theory, each with scale-dependent dressings. In this work, we consider the minimal truncation including up to four-fermion operators which permit diagnosing spontaneous symmetry breaking. The regulator function $R_k$ is not unique and can be chosen to optimize convergence and numerical stability. We employ the Litim regulator~\cite{Litim:2001up} which provides analytic handle on the flow equations. We refer technical details on the flow equation and truncation to \Cref{app:flows}. 

Functional approaches are widely used to study $\dSSB$, particularly in QCD and its phase structure, which are close in spirit to the theories studied here. Advanced fRG computations incorporate explicit momentum dependence in higher $n$-point functions, both in vacuum and at finite temperature or chemical potential, see e.g.~\cite{Fu:2022uow,Fu:2024ysj,Fu:2025hcm,Braun:2018bik,Braun:2019aow,Ihssen:2024miv}. Moreover, the fRG framework allows for scale-dependent field transformations along the flow, enabling the description of dynamically emerging effective degrees of freedom~\cite{Gies:2002hq,Fukushima:2021ctq,Pawlowski:2005xe}. The emergent composites (or dynamical hadronisation) approach facilitates the inclusion of higher-dimensional fermionic operators via bosonic interactions $\phi^n \propto\left(\bar\psi\,{\cal T} \psi\right)^{n}$, thus extending the analysis beyond the singularity of the four-fermion coupling that signals $\dSSB$ and providing smooth access to the chirally broken regime. This formulation offers both technical and conceptual advantages for studying strongly coupled gauge–fermion systems.

In this first work, we restrict ourselves to a qualitative analysis and leave the full bosonized treatment to future studies. Our goal is to identify which condensate forms at the highest scales, which is the first step towards the full IR dynamics. For this, it suffices to employ a minimal Fierz-complete truncation of four-fermion operators~\cite{Gies:2005as,Braun:2005uj,Braun:2006jd,Braun:2010qs}. 

The effective average action for generalized GG gauge-fermion theories has the form,
\begin{align}\label{eq:effaction}
\Gamma_k \left[A_\mu,\bar c ,c, \bar \psi,\psi, \bar \chi,\chi\right]=&\,\Gamma_{{\rm gauge},k}\left[A_\mu,\bar c ,c\right]+\Gamma_{{\rm Dirac},k}\left[A_\mu,\bar \psi,\psi, \bar \chi,\chi\right] +\Gamma_{{\rm 4F},k }\left[\bar \psi,\psi, \bar \chi,\chi\right] +\ldots\,
\end{align}
where $\ldots$ account for the remaining higher-dimensional operators which are present in the full effective action but are not taken into account in the minimal truncation here. The first two terms in \eqref{eq:effaction} stand for the pure gauge and Dirac parts including up to dimension four operators (e.g. dispersions and interaction terms). As mentioned, the information on $\dSSB$ and bound state formation is encoded in the purely fermionic operators which in \eqref{eq:effaction} constitute
\begin{align}\label{eq:effactionFermi}
\Gamma_{{\rm 4F},k }[\bar \psi,\psi,\bar \chi,\chi]= 
&-\int_x Z_{\psi}^2 \,  \sum_{i=1}^2  \lambda_{i} {\cal O}_i +Z_{\chi}^2 \sum_{i=3}^5  \lambda_i {\cal O}_i+Z_{\psi} Z_{\chi}\sum_{i=6}^7  \lambda_{i} {\cal O}_i \,,
\end{align}
and include all possible four-fermion operators forming a Fierz-complete basis which we construct with the Young Tensor Method~\cite{Li:2020gnx,Li:2020xlh,Li:2022tec}. In \eqref{eq:effactionFermi}, $Z_\psi$ and $Z_\chi$ are the fermionic field wave functions and $ \bar \lambda_{i}= k^{2} \; \lambda_{i} $ are the dimensionless form of the four-fermion dressings for the set of operators,
\begin{subequations}\label{eq:Fierzbasis}
\begin{align}
		{\cal O}_1 &= \big(\psi^\dag\bar{\sigma}^\mu  \psi\big)\big(\psi^\dag\bar{\sigma}^\mu  \psi\big),\\[1ex]
		{\cal O}_2 &= \big({\psi^\dag}^{f_1}\bar{\sigma}^\mu  \psi_{ f_2 }\big)\big({\psi^\dag}^{f_2}\bar{\sigma}^\mu  \psi_{ f_1}\big), \\[1ex]
		{\cal O}_3 & =  \big({\chi^\dag}^{ f_1}\bar{\sigma}^\mu  \chi_{f_2}\big)\big({\chi^\dag}^{f_2}\bar{\sigma}^\mu  \chi_{f_1}\big),\\[1ex]
		{\cal O}_4 & =  \big({\chi^\dag}\bar{\sigma}^\mu  \chi\big)\big({\chi^\dag}\bar{\sigma}^\mu  \chi\big),\\[1ex]
		{\cal O}_5 & =  \big({\chi^\dag}\bar{\sigma}^\mu \, T_{\rm anti}\,\chi\big)\big({\chi^\dag}\bar{\sigma}^\mu\,  T_{\rm anti} \,\chi\big),\\[1ex]
		{\cal O}_6 & =  \big(\psi^\dag\bar{\sigma}^\mu  \psi\big)\big({\chi^\dag}\bar{\sigma}^\mu  \chi\big),\\[1ex]
		{\cal O}_7 & =  \big(\psi^\dag\bar{\sigma}^\mu  \,T_{\rm a-fund}\,\psi \big)\big({\chi^\dag}\bar{\sigma}^\mu  \,T_{\rm anti} \,\chi \big)\,.
\end{align}
\end{subequations}
Here, we adopted the convention where fermionic color and the flavor ($f_{\ldots}$) indices are assumed contracted within each parenthesis unless explicitly displayed. $T_{\rm anti}$ and $T_{\rm a-fund}$ stand for the color group generators in the anti-symmetric and anti-fundamental representations, respectively. To construct this Fierz-complete basis we have imposed invariance under the Lorentz group, the gauge group $SU\left(N_c\right)$ and the whole flavor group $SU\left(\Ngen\left( N_c - 4\right)\right)\times SU\left(\Ngen\right) \times U\left(1\right)$, while taking into account the effect of repeated fields and Fierz redundancy, see \Cref{app:operatorconstruction} for details on the construction. 

The basis in \eqref{eq:Fierzbasis} includes all fermionic self-interaction terms at canonical dimension six, with the exception of the operator which appears only in the theory with $N_c=5$ and $N_{\rm gen}=1$:
\begin{align}\label{eq:tHooftoperator}
	{\cal O}_8 = \epsilon_{i_1 i_2 i_3 i_4 i_5} \left( \chi^{i_1 i_2}  \chi^{i_3 i_4} \right) \left( \chi^{i_5 j}  \psi_j \right),
\end{align}
where $\epsilon_{i_{\dots},\ldots}$ is the Levi-Civita tensor for the color group. 
This operator corresponds to the so-called ’t Hooft vertex, where the number of fermions is determined by the Dynkin indices of the fermions under the gauge group, which is associated with the axial anomaly and counts the number of fermion zero modes in the instanton background with unit topological charge. For theories with generic values of $\Ngen$ and $N_c$, the operator corresponding to the 't Hooft vertex involves $(N_c-4) \Ngen$ Weyl fermions of $\psi$ and $(N_c-2) \Ngen$ Weyl fermions of $\chi$, hence the operator has canonical mass dimension $\left[ {\cal O}_8 \right] = 3(N_c-3) \Ngen$. 

Since we focus on the large $\Ngen$ limit, we omit this operator in the present analysis as becomes increasingly canonically irrelevant. We further note that for the study of the single-generation case, the present operator basis further simplifies as ${\cal O}_1 \equiv {\cal O}_2$ and ${\cal O}_3 \equiv {\cal O}_4$.

\section{Patterns of symmetry breaking}\label{sec:phasediagram}
The evolution of four-fermion dressings is dictated by their respective RG flows, which are triggered purely by the gauge dynamics. While these flows have been extensively studied in QCD-like theories (see e.g.~\cite{Goertz:2024dnz,Dupuis:2020fhh,Fu:2022uow,Fu:2024ysj,Fu:2025hcm,Braun:2011pp}), as we will see, the situation in chiral-gauge theories with several species is qualitatively different and remains rather unexplored. We commence this Section by examining the general structure of the four-fermion RG flow in gauge–fermion systems and the underlying driving mechanism. We then discuss the resonant structure in this class of theories and provide more details about particular theories in \Cref{app:examples}.

\begin{figure*}[t!]
	\centering
    \includegraphics[width=.9\columnwidth]{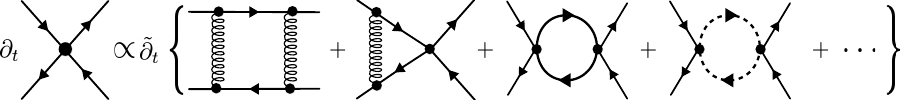}
	\caption{
	Sketch of the diagram topologies entering the RG flow of the four-$\psi$ point function relevant to the flow of the four-fermion couplings in \eqref{eq:flowlambda_structure}. Plain arrowed lines represent full $\psi$ propagators, dashed lines the respective $\chi$ propagators, and curly lines the full gauge boson propagators. The operator $\tilde\partial_t$ denotes a shorthand for the insertion of the regulators ($\partial_t R_k$) inherent to the flow equation \eqref{eq:floweq}, avoiding the need to write the different combinations of regulated lines explicitly. The $\ldots$ indicate other diagrams involving other higher dimensional operators (e.g. six-point functions), which are neglected in this analysis.
  	}
	\label{fig: RGflowdiagrams4F}
\end{figure*}

The RG flow of a general four-fermion coupling has the form,
\begin{align}
\partial_t \bar \lambda_{i} \propto& \,2\,\bar \lambda_{i}+ \boldsymbol{c}_{{\rm A},i} \cdot \alpha_g^2 +\boldsymbol{c}_{{\rm B},ij} \cdot \alpha_g\,\bar\lambda_{j} +\boldsymbol{c}_{{\rm C },ijk} \cdot \bar\lambda_{j} \bar\lambda_{k} +\ldots\,, \label{eq:flowlambda_structure}
\end{align}
as illustrated in the diagrammatic flow of the four $\psi$-point function sketched in \Cref{fig: RGflowdiagrams4F}. The first term in the right-hand side of \eqref{eq:flowlambda_structure} corresponds to the canonical scaling of the operator, and the following ones originate from vertex corrections and the fermion anomalous dimension. 
\begin{itemize}
\item The term $\alpha_g^2 = g^2/4\pi$ arises from purely gauge-mediated box diagrams (first diagram on the right-hand side of \Cref{fig: RGflowdiagrams4F}) and is encoded in the coefficient $\boldsymbol{c}_{{\rm A},i}$. The coefficients $\boldsymbol{c}_{{\rm A},i}$ act as constant shifts and therefore only change the vertical position of the flow without modifying its shape. While the four-fermion interactions are absent at the classical level, these are generated already at one loop by the box diagrams and therefore stress why fermion self-interactions should be interpreted as higher-order gauge corrections, with the various tensor structures corresponding to distinct interaction channels. 
\item Once induced, these interactions contribute to their own flow through the mixed term $\alpha_g \bar\lambda_j$, which represents a loop involving two gauge vertices and one four-fermion, accounted in $\boldsymbol{c}_{{\rm B},ij}$. The coefficients $\boldsymbol{c}_{{\rm B},ij}$ contribute as linear corrections act analogously to an anomalous dimension, i.e., as corrections to the canonical scaling.
\item Finally, the quadratic term $\bar\lambda_j \bar\lambda_k$, encoded in $\boldsymbol{c}_{{\rm C},ijk}$, describes purely fermionic self-interactions (right-most diagrams in \Cref{fig: RGflowdiagrams4F}). The indices $j,k$ run over all channels, including $i$, which corresponds to the basis in \eqref{eq:Fierzbasis}. 
\end{itemize}
The coefficients $\boldsymbol{c}_{{\rm A},i}$, $\boldsymbol{c}_{{\rm B},ij}$, and $\boldsymbol{c}_{{\rm C},ijk}$ encode the color and global symmetries of the theory and exhibit a non-trivial dependence on $N_c$ and $\Ngen$, as determined by traces over the internal structure of the theory.

Moreover, the $\ldots$ in \eqref{eq:flowlambda_structure} denote further subleading contributions involving one-loop diagrams with other higher dimensional operators. Given the one-loop structure of the flow equation \eqref{eq:floweq}, higher $n$-point functions (beyond six) can only contribute indirectly through lower ones. Additional terms with higher powers of $\alpha_g$ enter via the anomalous dimensions of the gauge and fermion fields, which appear through the regulator insertions. While these anomalous-dimension effects are included in the present approximation, higher-order corrections originating from six-point functions are not. For further details on this kind of flows, see~\cite{Braun:2011pp,Goertz:2024dnz,Dupuis:2020fhh} and references therein.

\begin{figure*}[t!]
	\centering
    \includegraphics[width=.6\columnwidth]{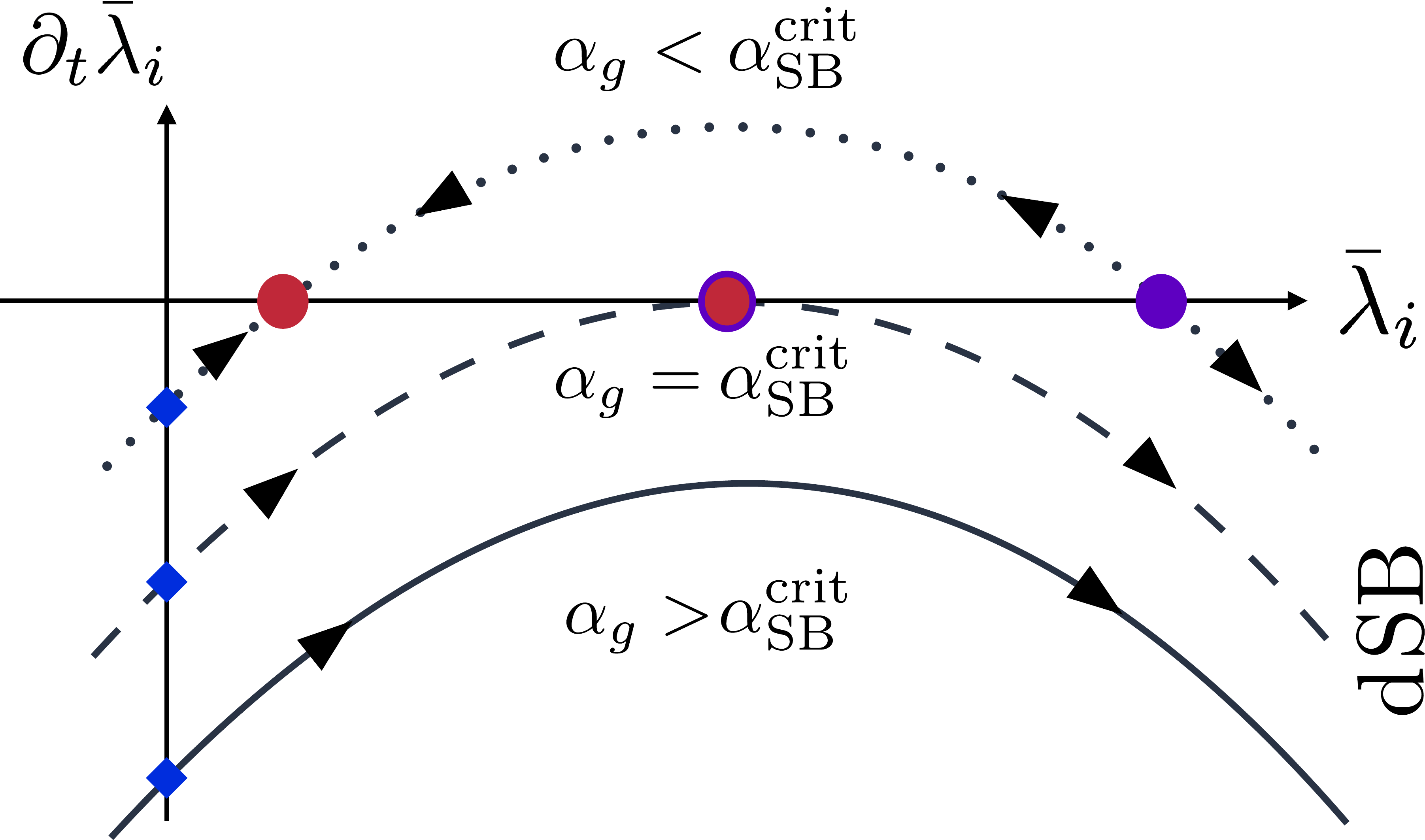}
	\caption{Depiction of the general form of the flow of a four-fermion coupling, which exhibits a resonant structure as a function of the coupling itself. The red and purple dots indicate an IR-attractive and a UV-repulsive fixed point, respectively. Increasing gauge corrections shift the flow downward, leading to a fixed-point merger at a critical strength $\acrit$. Beyond this point, the four-fermion coupling runs towards a singularity, signaling the onset of symmetry breaking.
  	}
	\label{fig:sketchRGflow}
\end{figure*}
The self-interaction terms $\boldsymbol{c}_{{\rm C},iii}$ in \eqref{eq:flowlambda_structure} shape the RG flow of a given four-fermion coupling as a function of itself into a parabola, see \Cref{fig:sketchRGflow,fig: flowlambda} for an explicit example. Depending on the sign of the self-interaction term (subject to convention), the parabola opens either upwards or downwards. For small $\alpha_g$, the flow of the couplings exhibits an IR-attractive fixed point, which in the classical limit ($k\to\infty$) reduces to the Gaussian one, $\alpha_g^*=0$ and $\bar\lambda_{i}^*=0$. This IR fixed point\footnote{In fact, a \textit{partial} fixed point of the full system as not all flows vanish (i.e. the gauge coupling flow in theories outside the conformal window) but only those respective to the four-fermion (or higher) interactions.} is dictated by the gauge dynamics and reflects once again the interpretation of four-fermion interactions as nothing else but higher-order gauge-fermion interactions. Importantly, the presence of such a fixed point bounds the maximum strength of the fermion self-interactions, thereby preventing the flow from developing a singularity diagnosing $\dSSB$.

Furthermore, as the gauge dynamics evolve towards the IR, the parabola shape is shifted upwards or downwards depending on the sign of the $\boldsymbol{c}_{{\rm A},i}$ coefficients, which read
\begin{align}\label{eq:cAcoefficients}
		\boldsymbol{c}_{{\rm A},\,i}&= \frac{9}{2} \bigg\{\;-\frac{1}{4 N_c^2}-\frac{3}{16},\;\,-\frac{N_c^2-8}{16 N_c},\;\,-1,\;\,-1 + \frac{ 4 + 2N_c}{N_c^2},\;\,1-\frac{N_c}{8}+\frac{4}{N_c},\;\, \notag \\[1.ex]&\hspace{7.5cm}-1 +\frac{N_c+2}{N_c^2},\;\,1-\frac{N_c}{4}+\frac{4}{N_c}\bigg\}\,.
\end{align}
Note that none of these depend on $\Ngen$, but only on $N_c$ (see also \Cref{fig:cAcoefficients} where we plot these coefficients). For a singularity to develop, the gauge dynamics must drive the disappearance of the IR fixed point, i.e., shift the RG flow parabola such that its zeros merge and vanish into the complex plane. This requires that the leading gauge corrections ($\boldsymbol{c}_{{\rm A},i}$) and the higher-order self-interactions ($\boldsymbol{c}_{{\rm C},iii}$) act coherently, namely by having the same sign, to induce a fixed-point annihilation.
This behavior is characteristic of a \textit{resonant} structure in the RG flow, cf.~\eqref{eq:flowlambda_structure} and \Cref{fig: RGflowdiagrams4F}, where the respective four-fermion coupling feeds back into its flow constructively. In the absence of such, the lower- and higher-order contributions act incoherently, thereby strengthening the persistence of the fixed point. While the alignment of $\boldsymbol{c}_{{\rm A},i}$ and $\boldsymbol{c}_{{\rm C},iii}$ having the same sign offers an accurate qualitative criterion for identifying a resonant structure, it assumes a simplified scenario where inter-channel contributions are subleading and the flow is dominated by the leading gauge terms. More intricate dynamics arise when off-diagonal terms become relevant, such as $\boldsymbol{c}_{{\rm B},ij}$ with $i \neq j$ or $\boldsymbol{c}_{{\rm C},ijk}$ with $i \neq j,k$, which can significantly alter the flow and the mechanism of symmetry breaking. These cases will be encountered below.

Moreover, the presence of a resonant structure in a given channel is a necessary but not sufficient condition for a singularity to occur, as the driving gauge dynamics must also reach a sufficient strength to cause the fixed-point merger. To quantify this, we define the critical coupling $\acrit$ as the minimal value of $\alpha_g$ for which at least one of the four-fermion flows ceases to exhibit an IR fixed-point, namely that  $\partial_t\bar\lambda_i< 0 $ or $\partial_t\bar\lambda_i>0 $ uniformly as a function of any $ \bar\lambda_j$. This can be summarized in a compact form in
\begin{align}\label{eq:alphacrit}
\acrit =\min \left\{ \alpha_g \,\middle|\, \exists\, i:\,\,\, \left(\, \partial_t\bar\lambda_i\leq 0 \quad \forall\, \bar\lambda_j \right) \,\,{\,\rm or \,}\,\left(\, \partial_t\bar\lambda_i\geq 0 \quad \forall\, \bar\lambda_j \right)\,\,\right\}\,.
\end{align}
It then follows, as depicted in \Cref{fig:sketchRGflow}, that for $\alpha_g < \acrit$, all four-fermion couplings $\bar\lambda_i$ flow to IR fixed points and for $\alpha_g > \acrit$, at least one coupling ceases to have a fixed point and flows towards a singularity, signaling the onset of $\dSSB$. The dominant channel is then identified as the one whose interaction term drives the flow away from the IR fixed point, thereby setting the dominant singularity of the system, that is, a channel whose omission would either eliminate the singularity or raise qualitatively the critical coupling. By contrast, subdominant channels can be removed without affecting the picture. For the class of theories considered here, the diagonal coefficients $\boldsymbol{c}_{{\rm C},iii}$ read:
\begin{align}\label{eq:cCcoefficients}
		\boldsymbol{c}_{{\rm C},\,iii}&=  \frac{1}{8\pi^2}\bigg\{\;N_c \Ngen\left( N_c - 4\right)-1,\; N_c+\Ngen\left( N_c - 4\right),\;2\,(N_c-1) N_c+4 \,\Ngen,\;\, \\[1.ex]
		&\hspace{1.5cm}  \frac{(N_c-1) N_c \Ngen}{2}-1,\, \frac{N_c \,(\Ngen+3)}{2}+\frac{8}{N_c}- \,\Ngen+1,\;-\frac{3}{2},\; \frac{N_c}{4}+\frac{3}{N_c}+\frac{3}{4}\bigg\}\,.\notag
\end{align}
Although these coefficients retain a constant sign for $N_c > 5$, the associated $\boldsymbol{c}_{{\rm A},i}$ may undergo a sign change (see \Cref{fig:cAcoefficients}). According to the resonance condition discussed above, this observation indicates that certain channels may become resonant in specific regions of theory space but not in others, thereby anticipating distinct channel dominance and symmetry-breaking patterns across different domains.

\begin{figure*}[t!]
	\centering
	\raisebox{0.25\height}{	\includegraphics[width=.45\columnwidth]{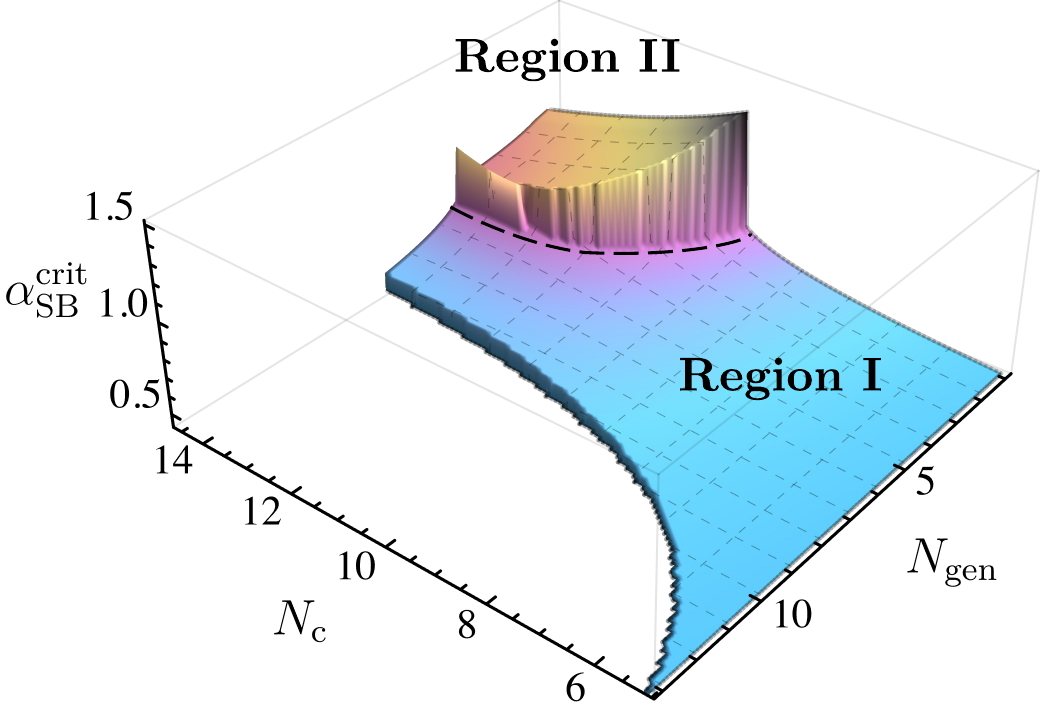}	}	\hspace{.5cm}
	\includegraphics[width=.45\columnwidth]{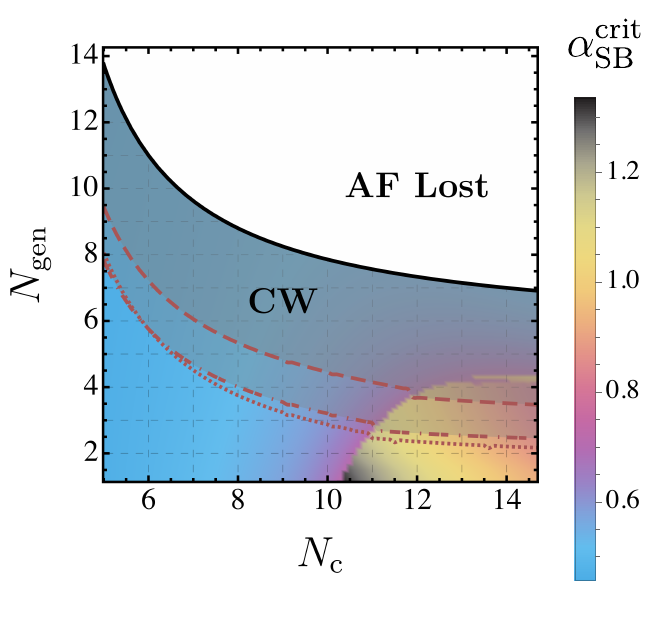}
	\caption{The three-dimensional plot on the left shows, as a colored surface, the critical strength of the gauge coupling ($\acrit$) required to trigger the dynamical generation of a scale in the space of generalized Georgi–Glashow theories, spanned by the number of colors ($N_c$) and generations ($N_{\rm gen}$). The right-hand plot displays the same surface projected onto the plane and different regimes are labeled. The solid black line marks the boundary of asymptotic freedom and the dashed, dashed-dotted and dotted pink lines indicate the theories where the $\acrit$ computed coincides with the IR fixed points from the two- (dashed), three- (dashed-dotted) and four-loop (dotted) $\overline{\text{MS}}$ beta functions, respectively. The shaded region above these curves shows the conjectured conformal window. 
  	}
	\label{fig: alphacrit}
\end{figure*}

To explore the $N_c-\Ngen$ landscape, we compute $\acrit$ by numerically solving the system of flow equations in \Cref{app:flows}. This analysis reveals the existence of distinct regions in theory space (see \Cref{fig: alphacrit}): 
\begin{itemize}[leftmargin=*]
	\item \textbf{{Region I:}} For $N_c \lesssim 11$, only the channels ${\cal O}_5$ and ${\cal O}_7$ exhibit the discussed naive resonant structure. Among these, ${\cal O}_5$ shows a clear dominance which means that its evolution is not affected by the other channels and that is the one which develops the singularity at the highest scale. Furthermore, the $\acrit$ condition is satisfied for the respective dressing first. The other channels have a negligible role in the dynamics and dropping them does not have an impact on the highest scale of symmetry breaking nor on the qualitative nature of the condensate. Moreover, as illustrated in \Cref{fig: alphacrit}, $\acrit$ remains essentially unchanged up to the vicinity of loss of asymptotic freedom. This result suggests that the leading condensate is $\langle \chi \chi \rangle$. Although this channel becomes singular at the highest scale and drives the initial symmetry-breaking dynamics, a more detailed IR analysis is required to determine the gauge representation of the condensate and the structure of its vacuum expectation value. Beyond this first scale, the next relevant interaction, ${\cal O}_7$, may give rise to a $\langle \psi \chi \rangle$ condensate. Interestingly, its resonant structure disappears around $N_c \sim 7$, potentially indicating the presence of subregions with qualitatively different IR behavior. Furthermore, the absence of resonance in the ${\cal O}_1$–${\cal O}_4$ channels suggests that $\langle \psi \psi \rangle$ condensates are unlikely to form. Nonetheless, to make definitive statements about the subsequent dynamical scales, a careful treatment of the decoupling of massive modes, both fermionic and gauge, is required.
\end{itemize}

As $N_c$ increases, the naive resonant structure in both ${\cal O}_5$ and ${\cal O}_7$ is lost due to the sign change in the associated $\boldsymbol{c}_{{\rm A},i}$'s. Around $N_c \sim 11$, the simple mechanism ceases to apply in any channel, yet a singularity still develops via a more involved process. For several generations ($N_{\rm gen} \gtrsim 5$ and $N_c \gtrsim 11$), off-diagonal contributions $\boldsymbol{c}_{{\rm B},5j}$ and $\boldsymbol{c}_{{\rm C},5jk}$  restore the resonant flow in $\bar \lambda_5$, effectively leading to the scenario in \textbf{Region I} but with a mild increase in $\acrit$. The fixed-point merger is now driven by higher-order terms (e.g. $\alpha_g\bar \lambda_j$ and $\bar \lambda_i \bar \lambda_j$) rather than the original $g^4$. This limit is more non-perturbative, and the key contributions arise from channels such as ${\cal O}_3$ and ${\cal O}_4$ feeding into ${\cal O}_5$. However, as will be discussed below, most of this region likely falls within the conformal window, as the gauge coupling required to trigger symmetry breaking is comparable to or exceeds the values estimated for the IR fixed points.

However, for large $N_c \gtrsim 11$ and small $\Ngen$, the off-diagonal contributions in the flow are too weak to induce a resonance in the previously dominant ${\cal O}_5$ channel. This transition reflects a change in the quantum realization of the gauge and global symmetries. Here we identify a further regime, whose precise boundaries depend sensitively on both $N_c$ and $\Ngen$:
\begin{itemize}[leftmargin=*]
	\item \textbf{{Region II:}} 
Here, all $\boldsymbol{c}_{{\rm A},i}$ are negative and therefore none of the channels fulfill the naive condition for the presence of a resonant structure. Nevertheless, we find a singularity. This arises from the interplay between different four-fermion channels in higher-order corrections, rather than being triggered by the primary gauge interactions. As a consequence, the critical coupling $\acrit$ is shifted to larger values, since higher-dimensional operators must first become large and IR-relevant. This points to a distinct non-perturbative mechanism underlying the initial scale generation which may be very different from the previously discussed and the present in QCD-like theories:
\begin{itemize}
    \item Given the large $\acrit$, it may indicate that multiple bifermion channels turn critical simultaneously and/or that multifermion condensates (encoded in even higher-dimensional operators) are generated.
    \item An alternative possibility is that color confinement sets in, generating a mass gap for the gauge modes, before $\acrit$ is reached. This seems plausible as in QCD-like theories the confining critical coupling is of the same order as the observed here \cite{Goertz:2024dnz}. In this case, the IR dynamics would be governed by confinement without the formation of fermion condensates, leading to a spectrum of massless baryons needed to match the global anomalies.  A specific realization of this idea has been shown to conflict with generalized anomaly matching \cite{Bolognesi:2019wfq,Bolognesi:2020mpe,Konishi:2024rjz}. However, more elaborate realizations, for instance involving distinct sets of massless baryonic states in the IR, could provide a consistent anomaly matching and still remain viable.
\end{itemize}
Disentangling such scenarios lies within the reach of functional methods \cite{Goertz:2024dnz,Ferreira:2025anh}, and a dedicated study of confinement dynamics in the GG model will be needed to establish whether a similar mechanism occurs there.
\end{itemize}

In summary, we found a rich landscape of global symmetry-breaking patterns which we summarized in \Cref{fig: alphacrit}. These regions include condensates that break the color group $SU(N_c)$ similarly to the diquark condensate in the color superconducting phase of QCD~\cite{Fukushima:2010bq,Alford:2007xm} and which lead to a dynamical Higgs phase with residual gauge dynamics below the first scale. The emergence of divergences and resonant channels depends sensitively on $N_c$, $N_{\rm gen}$, and the interplay of tensor structures. Identifying a dominant channel only indicates that a condensate within the fermion content of that channel is expected. More precisely, vacuum expectation values arise in specific fermion combinations organized into irreducible representations. For convenience, we chose a basis $\mathcal{O}_{1,\dots,7}$ in which the flow equations are explicit in $N_c$ and $N_{\rm gen}$, which limits a full treatment in terms of irreducible representations. Nevertheless, for particular values of $N_c$ and $N_{\rm gen}$ the linear change of basis can be computed, allowing a comparison in the irreducible basis. Even then, however, such an analysis does not fully capture the structure of the vacuum expectation values within the different irreducible condensates. Bosonization, by contrast, can clarify this structure, especially in regimes where higher-dimensional operators may be required to fully describe the dynamics. These results call for future work extending to larger $N_c$ and exploring exotic IR phenomena.

A further constraint remains to be addressed. Gauge–fermion theories are known to exhibit a conformal regime in the many-fermion limit, close to the boundary of asymptotic freedom. In this regime, a perturbative IR fixed point emerges, quantum scale invariance is realized, and no dynamical scale is generated.

By comparing $\acrit$ with the IR fixed-point values of the gauge beta function, one can estimate the lower boundary of the conformal window, following the approach used for QCD-like systems in~\cite{Goertz:2024dnz,Gies:2005as}. In fact, if the fixed point of  the gauge beta function is below the derived $\acrit$, then the gauge dynamics can never trigger spontaneous symmetry breaking, and eventually flows to an IR-interacting fixed point. In \Cref{fig: alphacrit}, we display the boundaries obtained with the IR fixed point of the two- (dashed), three- (dot-dashed), and four-loop (dotted) $\overline{\text{MS}}$ beta function. The four-loop coefficient, computed for the first time for this class of theories, is presented in \Cref{app:betafunctionsandFP}, along with further details relevant to this regime.

Last, given the recent interest in these models ~\cite{Bolognesi:2024bnm,Bolognesi:2020mpe,Bolognesi:2021yni,Bolognesi:2021jzs,Bolognesi:2019wfq,Bolognesi:2021hmg,Csaki:2021xhi,Tong:2021phe,Razamat:2020kyf,Cacciapaglia:2019vce}, particularly for $\Ngen = 1$, we now compare the fermion content and symmetry of the condensates. The most-attractive-channel (MAC) approach~\cite{Raby:1979my,Eichten:1981mu} evaluates the single-gluon exchange between two fermion lines in the irreducible representations and argues for the condensate to appear in the channel with strongest attraction. Our results always disfavor a $\langle \psi \psi \rangle$ condensate, consistent with the MAC. In \textbf{Region I}, we predict a leading $\langle \chi \chi \rangle$ condensate, while the MAC identifies $\langle \chi \chi \rangle$ for $N_c=5$, both $\langle \chi \chi \rangle$ and $\langle \psi \chi \rangle$ for $N_c=6$, and only $\langle \psi \chi \rangle$ for $N_c\geq7$. Nevertheless, the $\langle \psi \chi \rangle$ channel satisfies the naive resonant condition in part of \textbf{Region I}, suggesting that it may condense at lower energies. Approaches based on anomaly-matching arguments~\cite{Bolognesi:2024bnm,Bolognesi:2020mpe,Bolognesi:2021yni,Bolognesi:2021jzs,Bolognesi:2019wfq} also highlight the $\langle \psi \chi \rangle$ condensate in a color-flavor-locked vacuum. Remarkably, a $\langle \chi \chi \rangle$ condensate in these scenarios remains consistent with the anomaly conditions. Finally, the analysis from anomaly-mediated-Supersymmetry-breaking \cite{Csaki:2021xhi} predicts both $\langle \psi \chi \rangle$ and $\langle \psi \psi \rangle$ condensates, where the latter is manifestly disfavored in our findings.
\section{Conclusions}\label{sec:Conclusions}
In this work, we studied dynamical symmetry breaking in chiral-gauge theories, focusing on the emergence of fermion condensates that spontaneously break global and gauge symmetries, and consequently quantum scale invariance. Specifically, we have analyzed the class of generalized Georgi–Glashow theories, which contains two species of fermions transforming under different color representations and extended it to arbitrary numbers of generations to make contact to the loss of asymptotic freedom and a conjectured conformal window.

We constructed the Fierz-complete basis of four-fermion operators and derived their flows using the functional renormalization group. These carry the information on the channels which allow for a resonant growth and easily permit to compute the critical coupling necessary to trigger the dynamical generation of a scale. We find that the phase structure of generalized GG theories is rich as various regions with distinct patterns of $\dSSB$ exist. Such richness originates from the presence of multiple fermion species rather than from the chiral nature of the theory itself. This is as the latter imposes strong constraints on the allowed four-fermion operators, while the former introduces additional ones that mix different species. We note that, to the best of our knowledge, such type of multi-species gauge-fermion theories have not been studied before with approaches sensitive to the off-shell dynamics.

In \textbf{Region~I}, where $N_c\lesssim11$, the dominant tensor channel ${\cal O}_5$ becomes resonant, suggesting the formation of a $\langle \chi \chi \rangle$ condensate. However, for a larger number of colors in \textbf{Region~II}, the critical gauge coupling increases substantially, and the fixed-point merger is triggered by off-diagonal and higher-order contributions, suggesting a strongly coupled regime with qualitatively different underlying dynamics of non-perturbative nature. Furthermore, we provide an estimate on the boundary of the conjectured conformal window in the many generation limit, close to the loss of asymptotic freedom. This is obtained by comparing the derived critical coupling for dynamics to high-loop fixed-point values from perturbative $\beta$-functions, which the four-loop we derive in this work. 

Finally, we briefly comment on possible IR scenarios beyond the first condensate, where the interplay between different channels, the pattern of symmetry breaking and potential higher-order condensates remain open questions. A detailed resolution of these dynamics, including the emergence of color-flavor-locked phases or multi-fermion condensates, will require bosonized treatments or more elaborate truncations. 

The present analysis opens the door to many future studies. In an upcoming companion paper, we will extend this work in the symmetric case, commonly known as the Bars-Yankielowicz~\cite{Bars:1981se} model, exploring whether similar or qualitatively new symmetry-breaking patterns arise. Furthermore, the IR dynamics of these models can be tackled from first-principles in the present framework by dynamical bosonizing the dominant channel and including the dynamical appearance of a gauge-boson mass gap. More broadly, the functional method can be applied to investigate the mechanism of symmetric mass generation in various spacetime dimensions, see e.g.~\cite{Wang:2022ucy} for a review. Once all the 't Hooft anomalies are canceled, this mechanism can be used to generate a mass gap for chiral fermions without breaking the chiral symmetry. See e.g.~\cite{Wang:2013yta, Wang:2018ugf, Zeng:2022grc, Tong:2021phe} for the studies on the chiral fermion {3\,-\,4\,-\,5\,-\,0} model in 1+1 dimensions, and~\cite{Razamat:2020kyf, Tong:2021phe} for other related models in 3+1 dimensions.

\begin{acknowledgments}
	We would like to thank Jan M. Pawlowski for valuable comments and suggestions.  APG is supported by the RIKEN Special Postdoctoral Researcher (SPDR) Program. SV is supported by the Fonds de la Recherche Scientifique de Belgique (FNRS) under the IISN convention 4.4517.08. We are grateful to the Mainz Institute for Theoretical Physics (MITP) of the DFG Cluster of Excellence PRISMA+ (Project ID 39083149) for its hospitality and support during the initial stages of this work.
    H-.L.L. is supported by the start-up funding of Sun Yat-Sen University under grant number 74130-12255013 and by the National Science Foundation of China under Grants No.1250050417.
    The work of L.X.X. is partially supported by ``Exotic High Energy Phenomenology" (X-HEP), a project funded by the European Union - Grant Agreement n.101039756. This project is also supported by the Munich Institute for Astro-, Particle and BioPhysics (MIAPbP) which is funded by the Deutsche Forschungsgemeinschaft (DFG, German Research Foundation) under Germany's Excellence Strategy – EXC-2094 – 390783311.
    
\end{acknowledgments}

\appendix
\section*{Appendices}
\section{Four-fermion operator construction}\label{app:operatorconstruction}

This Appendix is devoted to the systematic construction of the four-fermion operators in generalized GG theories. We demand that operators are not only invariant under the $SU(N_c)$ gauge group, but also under the global flavor symmetry for each chiral fermion and the residual anomaly-free $U(1)$ symmetry as well.
This allows only three operator types: ${\chi^\dag}^2\chi^2$, ${\chi^\dag}\chi {\psi^\dag}\psi$ and ${\psi^\dag}^2\psi^2$, except for the scenario $N_c=5$ with $\Ngen=1$ in the GG model, where one additional operator type (and its complex conjugate) is allowed: $\psi\chi^3$ (see discussion in \ref{sec:dSSBandEA}.

For operator types ${\psi^\dag}^2\psi^2$ and ${\chi^\dag}^2\chi^2$, the repeated-fields effect needs to be taken into account the operator basis construction. The repeated fields are the ones with the same gauge and Lorentz quantum numbers, e.g. two $\psi$'s in the type ${\psi^\dag}^2\psi^2$. Due to the spin-statistics, operators should be totally anti-symmetric (symmetric) regarding to swapping any two of the repeated fields if the corresponding fields are fermions (bosons). 
Therefore, the permutation symmetry of the combined tensor of Lorentz, gauge and flavor should be totally anti-symmetric in these two types of operators. 
For ${\psi^\dag}^2\psi^2$, two independent invariant tensors can be used to contract the flavor indices:
\begin{align}
    T^{[f]}_{S,\psi} = \delta^{f_1}_{f_3}\delta^{f_2}_{f_4}+\delta^{f_1}_{f_4}\delta^{f_2}_{f_3},\\
    T^{[f]}_{A,\psi} = \delta^{f_1}_{f_3}\delta^{f_2}_{f_4}-\delta^{f_1}_{f_4}\delta^{f_2}_{f_3},
\end{align}
with $f_{1,2}$ the flavor indices for the field $\psi$ and $f_{3,4}$ for the field ${\psi^\dag}$, and the subscript $S,A$ indicates that the tensor is symmetric or anti-symmetric under the permutation of $1,2$ or $3,4$. 
Similarly, the gauge tensor can also be classified as $T^{[g]}_S$ and $T^{[g]}_A$ as follows:
\begin{align}
    T^{[g]}_{S,\psi} = \delta^{i_1}_{i_3}\delta^{i_2}_{i_4}+\delta^{i_1}_{i_4}\delta^{i_2}_{i_3},\\
    T^{[g]}_{A,\psi} = \delta^{i_1}_{i_3}\delta^{i_2}_{i_4}-\delta^{i_1}_{i_4}\delta^{i_2}_{i_3},
\end{align}
the $i$'s are gauge indices of the fields.
For the Lorentz tensor there is only one anti-symmetirc contraction: $T^{[L]}_{A}=\epsilon^{\alpha_1\alpha_2}\tilde{\epsilon}^{\dot{\alpha}_3\dot{\alpha}_4}$ with $\alpha_{1,2}$ and $\dot\alpha_{3,4}$ being the spinor indices of $\psi$ and ${\psi^\dag}$ respectively.
In order to construct a totally anti-symmetric tensor regarding to permuting two $\psi$ or $\psi^\dag$, $T^{[f]}_S$ ($T^{[f]}_A$) must be paired with $T^{[g]}_S$ ($T^{[g]}_A$), which results in two possible independent operators. To simplify our analysis we consider two independent linear combination of the operators, resulting in:
\begin{subequations}\label{eq:constructionpsi4operator}
\begin{align}
    {\cal O}^1 = ({\psi^\dag}^{f_1i_1}\bar{\sigma}^\mu \psi_{f_1 i_1})({\psi^\dag}^{f_2i_2}\bar{\sigma}^\mu \psi_{f_2 i_2}),\\
    {\cal O}^2 = ({\psi^\dag}^{f_1i_1}\bar{\sigma}^\mu \psi_{f_2 i_1})({\psi^\dag}^{f_2i_2}\bar{\sigma}^\mu \psi_{f_1 i_2}),
\end{align}
\end{subequations}
and the spinor indices in the parentheses are contracted implicitly.
The exact same reasoning can be applied to the type ${\chi^\dag}^2\chi^2$ as the Lorentz and flavor structures are the same as the type ${\psi^\dag}^2\psi^2$. It can be proven that the gauge invariant tensors can be organized into two symmetric  $T^{[g]}_{S,1/2,\chi}$ and one anti-symmetric $T^{[g]}_{A,\chi}$ for both antisymmetric and symmetric representations for $N_c\geq 4$. Therefore to form a totally anti-symmetric tensor, $T^{[g]}_{S,\{1,2\},\chi}$ should be paired with $T^{[f]}_{S\chi}$ and $T^{[g]}_{A,\chi}$
be paired with $T^{[f]}_{A,\chi}$, leading to three independent operators.
On the other hand, $T^{[g]}_{S,\{1,2\},\chi}$ and $T^{[g]}_{A,\chi}$ can be explicitly constructed with $\delta$ and generator $T^B$ in the following form:
\begin{align}
    T^{[g]}_{S,1,\chi} &= \delta^{a_1}_{a_3}\delta^{a_2}_{a_4}+\delta^{a_2}_{a_3}\delta^{a_1}_{a_4},\\
    T^{[g]}_{S,2,\chi} &= T^{B a_1}_{{\rm anti}\ a_3}T^{B a_2}_{{\rm anti}\ a_4}+T^{B a_2}_{{\rm anti}\ a_3}T^{B a_1}_{{\rm anti}\ a_4},\\
    T^{[g]}_{A,\chi} &= \delta^{a_1}_{a_3}\delta^{a_2}_{a_4}-\delta^{a_2}_{a_3}\delta^{a_1}_{a_4},
\end{align}
where $a_i$ are the gauge indices for $\chi$ or ${\chi^\dag}$, and repeated indices are contracted implicitly. From this equations, one can deduce that $\delta^{a_1}_{a_3}\delta^{a_2}_{a_4}$, $\delta^{a_2}_{a_3}\delta^{a_1}_{a_4}$ and $T^{B a_1}_{a_3}T^{B a_2}_{a_4}$ are also the three independent gauge tensors, which leads to the following three independent operators:
\begin{subequations}
\begin{align}
    {\cal O}^3 & =  ({\chi^\dag}^{f_1a_1}\bar{\sigma}^\mu \chi_{ f_1 a_2})({\chi^\dag}^{f_2a_2}\bar{\sigma}^\mu \chi_{ f_2 a_1}),\\
    {\cal O}^4 & =  ({\chi^\dag}^{f_1a_1}\bar{\sigma}^\mu \chi_{ f_1 a_1})({\chi^\dag}^{f_2a_2}\bar{\sigma}^\mu \chi_{ f_2 a_2}),\\
    {\cal O}^5 & =  ({\chi^\dag}^{f_1a_1}\bar{\sigma}^\mu T^{Ba_2}_{a_1}\chi_{ f_1 a_2})({\chi^\dag}^{f_2a_3}\bar{\sigma}^\mu T^{Ba_4}_{a_3}\chi_{f_2 a_4}).
\end{align}
\end{subequations}
For the operator of the type ${\chi^\dag}\chi {\psi^\dag}\psi$, it is quite obvious that there are two ways to form gauge singlet -- either with $\delta$'s or generators $T$ to contract two pairs of conjugate fields. Flavor-wise there is only one way to contract the indices, leading to 2 independent operators:
\begin{subequations}
\begin{align}
        {\cal O}^6 & =  ({\psi^\dag}^{f_1i_1}\bar{\sigma}^\mu \psi_{ f_1 i_1})({\chi^\dag}^{f_2a_2}\bar{\sigma}^\mu \chi_{ f_2 a_2}),\\
    {\cal O}^7 & =  ({\psi^\dag}^{f_1i_1}\bar{\sigma}^\mu T^{Bi_2}_{i_1}\psi_{ f_1 i_2})({\chi^\dag}^{f_2a_3}\bar{\sigma}^\mu T^{Ba_4}_{a_3}\chi_{f_2 a_4}). 
\end{align}
\end{subequations}
In the above notations, we write the generator of anti-symmetric representation as a matrix in ordinary way with two indices for the anti-symmetric representation. However, one can also use the notation of ~\ref{eq:tHooftoperator}, where the field $\chi$ is written in terms of two anti-symmetric fundamental indices. In this notation, one can write the generator $T_{\rm anti}$ in terms of generator of fundamental representation and the $\delta$-function that contract a fundamental and an anti-fundamental indices. Formally, it can be expressed as follows:
\begin{align}
    (T^A_{\rm anti})_{(i_1,j_1)}{}^{({i}_2,{j}_2)}=\delta^{{j_2}}_{j_1}T_{fi_1}^{A{i}_2}-\delta^{{i}_1}_{j_1}T_{fi_1}^{A{j}_2}
    -\delta^{{j}_2}_{i_1}T_{fj_1}^{A{i}_2}+\delta^{{i}_2}_{i_1}T_{fj_1}^{A{i}_2}\;,
\end{align}
where $T_f$ is the generator of the fundamental representation. 
The advantage of using this notation is that in the computation of the flow equation, the trace involving the antisymmetric generators can always be converted to the trace of the fundamental generators, which can be simplified with programs such as \texttt{FeynCalc}~\cite{Shtabovenko:2023idz}.

\section{Full effective action and flow equations}\label{app:flows}
We proceed providing the flows of the dressing functions for all four-fermion operators previously constructed. These are obtained directly from the diagrammatic flow of purely fermionic four-point functions, which can be systematically and consistently evaluated using automated symbolic tools~\cite{Huber:2019dkb,Pawlowski:2021tkk}. The computation follows closely the derived in~\cite{Gies:2005as,Goertz:2024dnz}, and we now review it in detail for the particular case of the pure-$\psi$ operators.

The diagrammatic flow of the four-$\psi$ vertex reads: 
\begin{align}\label{eq:flow4PFpsi}
 \partial_t \Gamma_k ^{(\bar\psi_a \psi_b \bar\psi_c\psi_d)}= -\partial_t\left(Z^2_\psi \,\lambda_1 \right) {\cal T}^{abcd}_{1,L} - \partial_t\left(Z^2_\psi  \lambda_2 \right) {\cal T}_{2,L}^{abcd}\,,
\end{align}
which results from taking four functional derivatives of the Wetterich equation with respect to fermionic fields. Here ${\cal T}^{\ldots}_{1/2}$ are the tensors structures constructed in \eqref{eq:constructionpsi4operator} which contain gauge/flavor/Lorentz indices for the respective operators ${\cal O}_{1/2}$. To disentangle the dressing functions associated to the different tensor structures, one must employ appropriate projections to contract the open indices. 

The most efficient strategy is to construct projection operators such that each projected correlator depends solely on one of the dressing functions~\cite{Braun:2025gvq,Gehring:2015vja}. This is in fact, particularly relevant when bosonizing the theory. A simpler alternative, sufficient for the present purpose, is to construct a complete basis of projectors directly adapted to the tensor structures under consideration. We define:
\begin{align}
	{\cal P}_{i,R}^{abcd} = {\cal T}_{i,R}^{badc}\,,
\end{align}
as the projector which, when applied to~\eqref{eq:flow4PFpsi}, yield the projected flow equations:
\begin{subequations}\label{eq:projectedflows}
\begin{align}
\partial_t  \left[ {\cal P}_{1,R}^{abcd} \Gamma_k^{(\bar\psi_a \psi_b \bar\psi_c\psi_d)} \right] &= - \partial_t\left(Z^2_\psi \lambda_1 \right) \left[ {\cal P}_{1,R}^{abcd} {\cal T}_{1,L}^{abcd} \right] 
- \partial_t\left(Z^2_\psi \lambda_2 \right) \left[ {\cal P}_{1,R}^{abcd} {\cal T}_{2,L}^{abcd} \right]\,, \\[1ex]
\partial_t  \left[ {\cal P}_{2,R}^{abcd} \Gamma_k^{(\bar\psi_a \psi_b \bar\psi_c\psi_d)} \right] &= - \partial_t\left(Z^2_\psi \lambda_1 \right) \left[ {\cal P}_{2,R}^{abcd} {\cal T}_{1,L}^{abcd} \right] 
- \partial_t\left(Z^2_\psi \lambda_2 \right) \left[ {\cal P}_{2,R}^{abcd} {\cal T}_{2,L}^{abcd} \right]\,.
\end{align}
\end{subequations}
The $R$ subindex stands for a replacement of $\bar\sigma^\mu\to \sigma^\mu$ in the respective chiral tensor structure.
Solving \eqref{eq:projectedflows} yields the flows of the dressing functions $\bar\lambda_1$ and $\bar\lambda_2$:
\begin{subequations}
\begin{align}
\partial_t \bar\lambda_1 &= (2 + 2 \eta_\psi) \bar\lambda_1  \\[1ex]
 & \;\; + \frac{k^2}{Z_\psi^2} \frac{ 
\partial_t \left[ {\cal P}_{1,R}^{abcd} \Gamma_k^{(\bar\psi_a \psi_b \bar\psi_c\psi_d)} \right] (N_c N_\psi + 1)  \notag
- \partial_t \left[ {\cal P}_{2,R}^{abcd} \Gamma_k^{(\bar\psi_a \psi_b \bar\psi_c\psi_d)} \right] (N_c + N_\psi)
}{128 N_c (N_c^2 - 1) N_\psi (N_\psi^2 - 1)}\,, \\[2ex]
\partial_t \bar\lambda_2 &= (2 + 2 \eta_\psi) \bar\lambda_2  \\[1ex]
&\;\;+ \frac{k^2}{Z_\psi^2} \frac{ 
\partial_t \left[ {\cal P}_{2,R}^{abcd} \Gamma_k^{(\bar\psi_a \psi_b \bar\psi_c\psi_d)} \right] (N_c N_\psi + 1) 
- \partial_t \left[ {\cal P}_{1,R}^{abcd} \Gamma_k^{(\bar\psi_a \psi_b \bar\psi_c\psi_d)} \right] (N_c + N_\psi)
}{128 N_c (N_c^2 - 1) N_\psi (N_\psi^2 - 1)}\,. \notag
\end{align}
\end{subequations}
This procedure can be repeated for all other four-fermion operators to obtain their respective flows of the dressings. 

The diagrammatic flows are computed in the vanishing external momentum limit, which, together with the choice of the Litim regulator grants us access to the following analytic expressions for the flows:
\begin{subequations}
\label{eq:flowslambda}
\begin{align}
\partial_t \bar \lambda_1 =& (2+2\, \eta_\psi) \bar\lambda_1 \notag \\[1ex] 
&+ \frac{1}{16 \pi^2}\bigg[ \frac{g^4 \left(3 N_c^2+4\right) (5 \eta_A+3 \eta_\psi-45)}{160 N_c^2}+\frac{g^2 (5 \eta_A+6 \eta_\psi-60) (\bar \lambda_1-\bar \lambda_2 N_c)}{10 N_c} \notag \\[1ex] 
&-\frac{1}{20}(\eta_\chi-5) N_\chi \left( \bar \lambda_6^2 (N_c-1) N_c +\frac{ (N_c-2) }{2 N_c}\bar \lambda_7^2 \right) \notag \\[1ex] 
&+\frac{4}{5} (\eta_\psi-5) \left( \bar \lambda_2^2- \bar \lambda_1 \bar \lambda_2 (N_c+N_\psi)- \bar \lambda_1^2\frac{(N_c N_\psi-1)}{2}\right)
\bigg]\,,\\
\partial_t \bar \lambda_2 =& (2+2\, \eta_\psi) \bar\lambda_2 \notag \\[1ex] 
&+ \frac{1}{16 \pi^2}\bigg[ \frac{g^4 \left(N_c^2-8\right) (5 \eta_A+3 \eta_\psi-45)}{160 N_c}+\frac{g^2 (5 \eta_A+6 \eta_\psi-60) (\bar \lambda_2-\bar \lambda_1 N_c)}{10 N_c}\notag\\
&-\frac{1}{40} (\eta_\chi-5) \bar \lambda_7^2 (N_c-2) N_\chi+\frac{2}{5} (\eta_\psi-5) \left( 4 \bar \lambda_1 \bar \lambda_2- \bar \lambda_2^2 (N_c+N_\psi) \right)
\bigg]\,,
\end{align}
\begin{align}
\partial_t \bar\lambda_{3} =& (2+2\, \eta_\chi) \bar\lambda_3 \notag \\[1ex] 
& + \frac{1}{16 \pi^2}\bigg[ \frac{1}{10} g^4 (5 \eta_A+3 \eta_\chi-45)+\frac{ \left(g^2(5 \eta_A+6 \eta_\chi-60)\left( N_c \bar \lambda_5-2\bar \lambda_3 \right) \right)}{5 N_c}\notag\\
&+\frac{ \eta_\chi-5}{5}  \left( 8\bar \lambda_4 \bar \lambda_3 + \frac{4 \left(N_c^2-N_c+4\right)}{N_c} \bar \lambda_5 \bar \lambda_3 - \bar \lambda_3^2 \left(N_c^2-N_c+2 N_\chi\right)+  6 \bar \lambda_5^2 \right)
\bigg]\,,\\
	\partial_t \bar\lambda_4 =& (2+2\, \eta_\chi) \bar\lambda_4 \notag \\[1ex]&+ \frac{1}{16 \pi^2}\bigg[ \frac{g^4 \left(N_c^2-2 N_c-4\right) (5 \eta_A+3 \eta_\chi-45)}{10 N_c^2}\notag\\
    & -g^2\frac{\left(5 \eta _A+6 \eta _{\chi }-60\right) \left(\lambda _5 \left(N^2{}_c-2 N_c-4\right)+2 \lambda _3 N_c\right)}{5 N_c^2} \notag \\ \notag
	&+ \frac{\left(\eta _{\chi }-5\right)}{10}  \left( \frac{48 \bar \lambda _5 \bar  \lambda _3}{N_c}+\frac{\bar  \lambda _5^2 \left(12 \left(N_c^2-2 N_c-4\right)\right)}{N_c^2}+\frac{\bar \lambda _4 \bar \lambda _5 \left(8 \left(-N_c^2+N_c+2\right)\right)}{N_c} \right) \notag \\
    & + \frac{\left(\eta _{\chi }-5\right)}{10} \left(-4 \bar  \lambda _4 \bar \lambda _3 \left(N_c^2-N_c+2 N_\chi \right)+\bar \lambda _4^2 \left(-2  N_\chi N_c^2+2  N_\chi N_c+4\right)+8 \bar  \lambda _3^2\right) \notag \\[1ex]& -\frac{\left(\eta _{\psi }-5\right)}{10} N_\psi N_c \bar \lambda_6^2 \bigg]\,,
\end{align}
\begin{align}
\partial_t\bar \lambda_{5} =& (2+2\, \eta_\chi) \bar\lambda_5 \notag \\[1ex]&+ \frac{1}{16 \pi^2}\bigg[ \frac{g^4 \left(N_c^2-8 N_c-32\right) (5 \eta_A+3 \eta_\chi-45)}{80 N_c}\notag \notag \\[1ex]&+\frac{g^2 (5 \eta_A+6 \eta_\chi-60) (\bar \lambda_5 (N_c+4)-N_c (\bar \lambda_4+\bar \lambda_3))}{5 N_c}\notag\\
&+\frac{1}{5}  (\eta_\chi-5) \left( 8 \bar \lambda_4 \bar \lambda_5 -4 \bar \lambda_3 \bar \lambda_5 (N_\chi-3)-\frac{  \left(N_c^2 (N_\chi+3)-2 N_c (N_\chi-1)+16\right)}{ N_c} \bar \lambda_5^2 \right) \notag\notag \\[1ex]&-\frac{1}{20} (\eta_\psi-5) \bar \lambda_7^2 N_\psi
\bigg]\,,
\end{align}
\begin{align}
\partial_t \bar\lambda_{6} =& (2+\, \eta_\psi+\, \eta_\chi) \bar\lambda_6 \notag \\[1ex]& + \frac{1}{16 \pi^2}\bigg[ \frac{g^4 \left(N_c^2-N_c-2\right) (10 \eta_A+3 (\eta_\chi+\eta_\psi-30))}{20 N_c^2}\notag\\
&-\frac{g^2 \bar \lambda_7 \left(N_c^2-N_c-2\right) (5 \eta_A+3 (\eta_\chi+\eta_\psi-20))}{10 N_c^2}\notag  \\[1ex]&+\frac{3 (\eta_\chi+\eta_\psi-10)}{10} \left( \bar \lambda_6^2 +\frac{\left(N_c^2-N_c-2\right) }{2 N_c^2} \bar \lambda_7^2  \right)\notag\\
&-\frac{1}{5} (\eta_\chi-5) \Bigg( \bar \lambda_4 \bar \lambda_6 \left(N_c^2 N_\chi-N_c N_\chi+2\right)+ \bar \lambda_3 \bar \lambda_6 \left(N_c^2-N_c+2 N_\chi\right)  \notag \\[1ex]&\hspace{7cm}+  \frac{2 \left(N_c^2-N_c-2\right)}{N_c} \bar \lambda_5 \bar \lambda_6  \Bigg) \notag\\
&-\frac{2}{5} (\eta_\psi-5) \left( \bar \lambda_1 \bar \lambda_6 (N_c N_\psi+1)+\bar \lambda_2 \bar \lambda_6 (N_c+N_\psi) \right) \bigg]\,,\\
\partial_t \bar \lambda_{7} =& (2+\, \eta_\psi+\, \eta_\chi) \bar\lambda_7 \notag \\[1ex]&+ \frac{1}{16 \pi^2}\bigg[ \frac{g^4 \left(N_c^2-4 N_c-16\right) (10 \eta_A+3 (\eta_\chi+\eta_\psi-30))}{80 N_c}\notag\\
&+\frac{g^2 (5 \eta_A+3 (\eta_\chi+\eta_\psi-20)) (\bar \lambda_7 (N_c+4)-2 \bar \lambda_6 N_c)}{10 N_c}  \notag\\
&+\frac{(\eta_\chi+\eta_\psi-10)}{5}  \left( 3 \bar \lambda_6 \bar \lambda_7 -\frac{ \left(N_c^2+3 N_c+12\right) }{4 N_c}\bar \lambda_7^2 \right)\notag \\[1ex]&-\frac{2(\eta_\psi-5)}{5} \left( \bar \lambda_1 \bar \lambda_7+ \bar \lambda_2 \bar \lambda_7 N_\psi\right)\notag\\
&-\frac{2(\eta_\chi-5)}{5}   \left( \bar \lambda_4 \bar \lambda_7+\bar \lambda_3 \bar \lambda_7 N_\chi  -\frac{  \left(N_c^2 (N_\chi+1)-2 N_c (N_\chi+1)-4\right)}{10 N_c} \bar \lambda_5 \bar \lambda_7 \right)
\Bigg]\,.
\end{align}
\end{subequations}
Here, $\eta_A$, $\eta_\psi$ and $\eta_\chi$ stand for the anomalous dimensions of the gauge and fermion fields respectively. These are computed from the respective two-point functions
\begin{align}
    \eta_A =& -\frac{\partial_t Z_A}{Z_A}= -\left.\frac{\partial_{p^2}\,{\rm tr}\left[ {\cal P}^{(A A)} \partial_t\Gamma^{(AA)}_k\right]}{Z_{A}\,{\rm tr}\left[{\left({\cal P}^{(A A)}\right)}^2\right] }\right|_{p=0}=\frac{g^2}{16\pi^2}\bigg[\left(\frac{31}{36} \eta_A-\frac{11}{3}\right)  N_c  \notag \\[1ex]&\hspace{7cm}-\frac{(\eta_\chi-4)}{6}   (N_c-2) N_\chi-\frac{(\eta_\psi-4)}{6}   N_\psi\bigg]\,,\\
    \eta_\psi =& -\frac{\partial_t Z_\psi}{Z_\psi}= -\left.\frac{\partial_{p^2}\,{\rm tr}\left[ {\cal P}^{(\psi\psi)} \partial_t\Gamma^{(\bar \psi\psi)}_k\right]}{Z_{\psi}\,{\rm tr}\left[{\left({\cal P}^{(\psi\psi)}\right)}^2\right] }\right|_{p=0}=\frac{g^2\left(N_c^2-1\right)}{32 \pi^2N_c}\left[\frac{(\eta_\psi-4) }{4 }-\frac{(\eta_A-5)  }{5 }\right]\,,\\
    \eta_\chi =& -\frac{\partial_t Z_\chi}{Z_\chi}= -\left.\frac{\partial_{p^2}\,{\rm tr}\left[ {\cal P}^{(\chi\chi)} \partial_t\Gamma^{(\bar \chi\chi)}_k\right]}{Z_{\chi}\,{\rm tr}\left[{\left({\cal P}^{(\chi\chi)}\right)}^2\right] }\right|_{p=0}=\frac{ g^2\left(N_c^2-N_c-2\right)}{16 \pi^2 N_c}\left[\frac{(\eta_\chi-4) }{4 }-\frac{(\eta_A-5)  }{5 }\right]\,,
\label{eq:etaAkUV}
\end{align}
and given that the anomalous dimensions enter in the regulated lines they  lead to a closed system of equations which can be solved giving infinite-order expressions in a resumed form. This feature manifests one of the sources of the non-perturbative character of this fRG approach. The  $\cal P^{(\ldots)}$ stand for the projectors onto the kinetic terms of the respective fields. 

Furthermore, from the flows in \eqref{eq:flowslambda}, the previously discussed coefficients ($\boldsymbol{c}_{\ldots}$) can be read off as a function of the number of chiral matter. To facilitate the identification of each contribution, we have displayed the flows as a function of the fermion multiplicities given by the global symmetries,
\begin{align}
 	G_f=SU\left(N_\psi\right)\times SU\left(N_\chi\right) \times U\left(1\right)\; ,
\end{align}
where  the anomaly-free combination fixes,
\begin{align}
 	&N_\psi=\Ngen\left( N_c - 4\right) &&{\rm and } &&N_\chi=\Ngen\,.
\end{align}

So far, we have not specified the pure gauge and Dirac parts of the action. In this work we employ a truncation in this sectors which includes the interaction terms up to dimension four and the classical tensor structures which is the same one as in \cite{Goertz:2024dnz}. Furthermore, in the gauge fixing sector we have used the Landau gauge. As an additional cross-check, we have derived the flow of the gauge couplings from all the included avatars and obtained the correct one-loop structure.

\begin{figure}[t!]
	\centering
	\includegraphics[width=0.75\columnwidth]{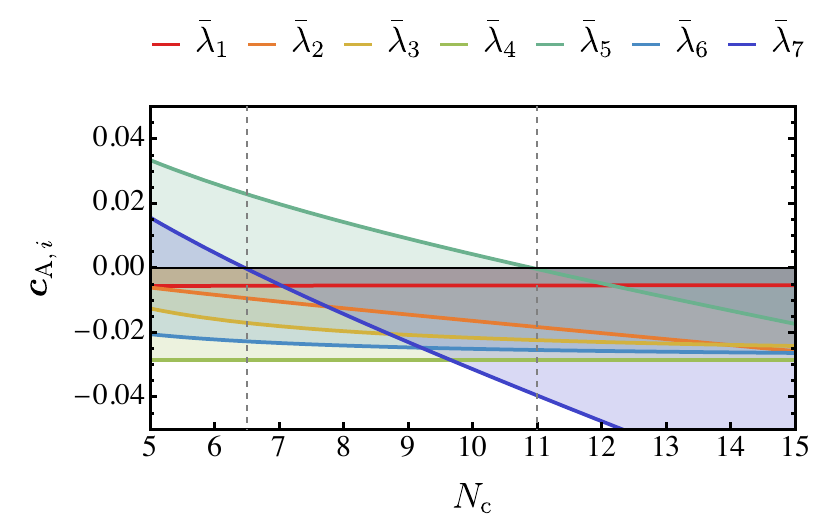}	
	\caption{ Value of the $\boldsymbol{c}_{{\rm A},i}$ coefficients which reflect the strength of the primer gauge corrections on the generation of the different four-fermion dressings. The sign crossing  of $\boldsymbol{c}_{{\rm A},5}$  at $N_c\sim 11$ and $\boldsymbol{c}_{{\rm A},7}$ provides the identification of regimes and potential subdomains.}
	\label{fig:cAcoefficients}
\end{figure}

Some further remarks are in order. First, note that only  $\boldsymbol{c}_{{\rm C},ijk}$ depends on $\Ngen$, because the only source of $\Ngen$ dependence in the flow arises from pure fermionic loops, which contribute exclusively to $\bar \lambda_i \bar\lambda_j$ terms.

Moreover, part of the system has a resemblance to QCD which permits to perform some cross checks. Operators ${\cal O}_1$ and  ${\cal O}_2$ are equivalent to those known as  (V-A) and (V-A)$_{\rm adj}$ but here only with one chirality \cite{Gies:2005as,Braun:2011pp,Goertz:2024dnz}. In fact, we find an agreement up to a factor 1/2 in the $N_\psi$ contributions in their respective flows, providing a consistency check to our computation. 

Furthermore, this comparison also allows us to learn more details about the structure of gauge-fermion theories. The computation here presented also allows us to understand the transformation of QCD like results to the chiral limit. One may start with a Dirac QCD-like basis \cite{Gies:2005as,Braun:2011pp,Goertz:2024dnz} and decouple one of the chiralities. Note that this is merely useful to understand the structural level as the product theories are anomalous.
The first consequence is that the scalar-pseudoscalar and vector+axial channels are not generated at all as required the presence of two chiralities, as the diagrams are all proportional to $N_L N_R$. As seen here, in the QCD-like scenario the ${\cal O}_{({\rm V-A})}$ and ${\cal T}_{({\rm V-A})_{{\rm adj}}}$ channels (analogues to our ${\cal O}_1$ and ${\cal T}_2$) do not fulfill the lowest-order structure for triggering a resonance. Nonetheless, as discussed in \cite{Goertz:2024dnz},  in the many-flavor QCD limit with locking dynamics evidences for a further resonant ${\cal T}_{({\rm V-A})_{{\rm adj}}}$ channels and condensates has been found, mainly led by strong fermion dynamics in the (S-P) channel. Moreover, the remaining vector-axial structure is indeed generated leading to the results here obtained. A last remark is in order, in the presence of one single chirality the charge-parity symmetry is not present and terms that would violate it in vector-like theories can appear. For example, the four-fermion interactions now receive contributions of the form
\begin{align}\label{eq:CPmomentum}
	(N_L-N_R)\,\,	\epsilon^{\mu\nu\sigma\rho}p_1^\mu p_2^\nu p_3^\sigma q^\rho
\end{align}
which in the QCD limit exactly vanish but here preserve. The $p_i$'s stand for the external momenta of the four-point function which, in the present vanishing external-momenta approximation such contributions drop. Nonetheless, we highlight this type of novel higher-order corrections encoded in the momentum dependence of the flows.

\section{A closer look at particular examples}\label{app:examples}

To better illustrate the underlying dynamics and the analysis performed in this work, this appendix presents a detailed discussion of the integrated RG flows for selected representative theories. We simultaneously integrate the flows in~\eqref{eq:flowslambda} along with the flow of the gauge coupling. The four-fermion flows incorporate the resumed form of the anomalous dimensions, which effectively accounts for infinite-order corrections.

As previously discussed, we derived the RG flow for various avatars of the gauge coupling and confirmed that the one-loop structure is the correct one. Nevertheless, for the purposes of this appendix, we employ the multi-loop perturbative RG equations, which will be presented in \Cref{app:betafunctionsandFP}. This choice implies that the four-fermion couplings do not feed back self-consistently into the evolution of the gauge coupling. However, the equivalent higher-order effects are already encoded in the perturbative running. While performing the fully self-consistent integration poses no technical difficulty, for the qualitative discussion at hand this simplification introduces no significant differences. In fact, using the perturbative expression allows us to correctly implement the multi-loop symmetry factors, which are relevant for fixed-point determination and for connecting to the conformal regime.

We emphasize that the analysis in the main text does not rely on this approximation. The determination of $\acrit$ depends solely on the flow of the four-fermion dressings, which are only affected by the truncation through the gauge-field anomalous dimension. Note that we are not attempting to capture the full IR dynamics, and in particular we have neglected confinement and the formation of a mass gap in the gauge sector.

\begin{figure*}[t!]
	\centering
	\includegraphics[width=.55\textwidth]{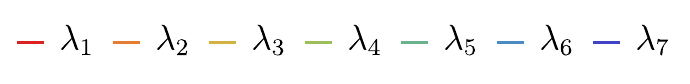}\\
	\begin{minipage}{0.45\textwidth}
		\includegraphics[width=\columnwidth]{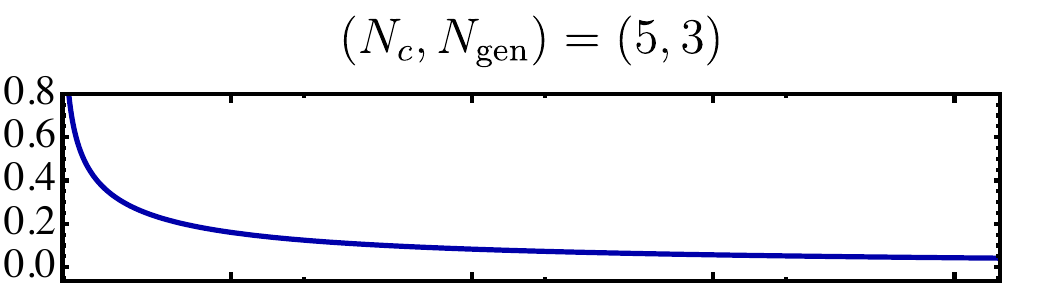}\vspace{-0.cm}
		\includegraphics[width=\columnwidth]{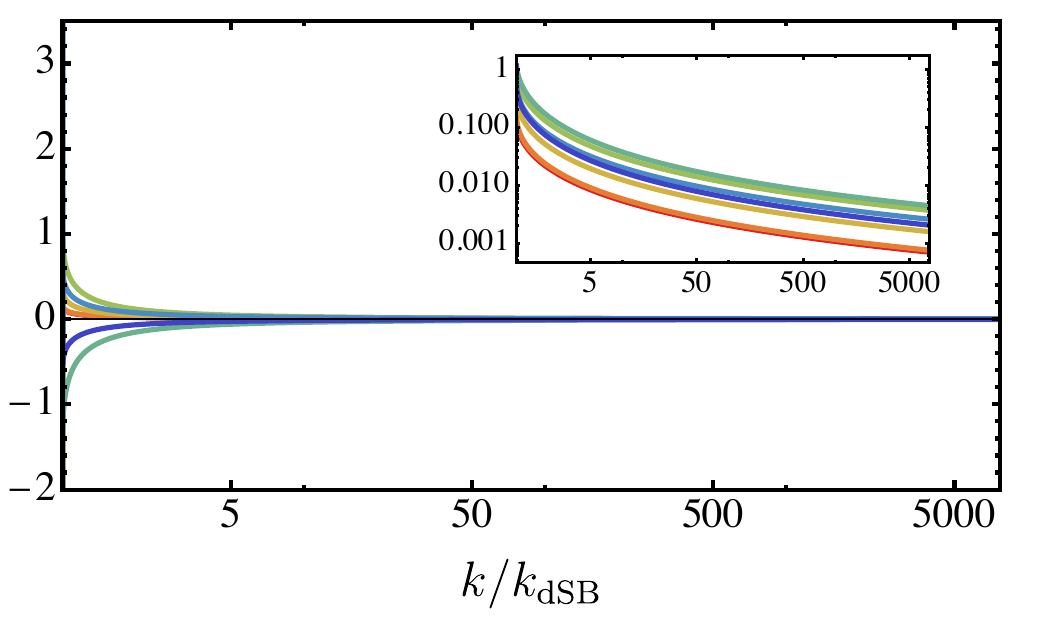}	
	\end{minipage}
	\begin{minipage}{0.45\textwidth}
		\includegraphics[width=\columnwidth]{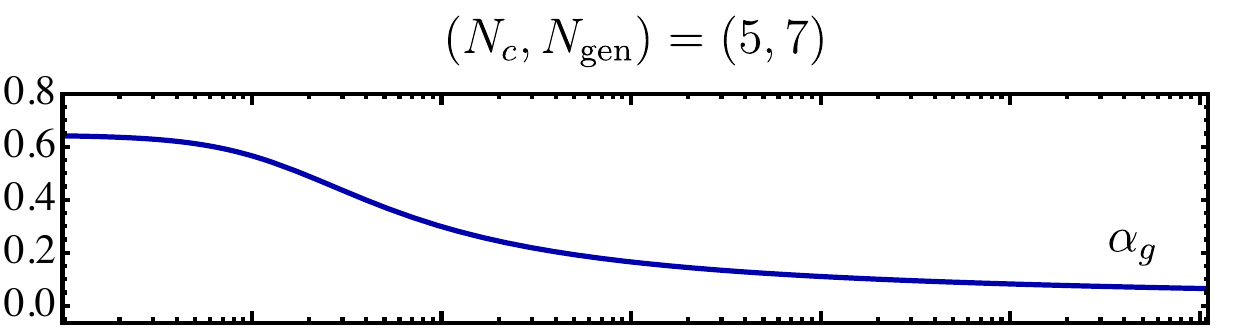}\vspace{-0.cm}
		\includegraphics[width=\columnwidth]{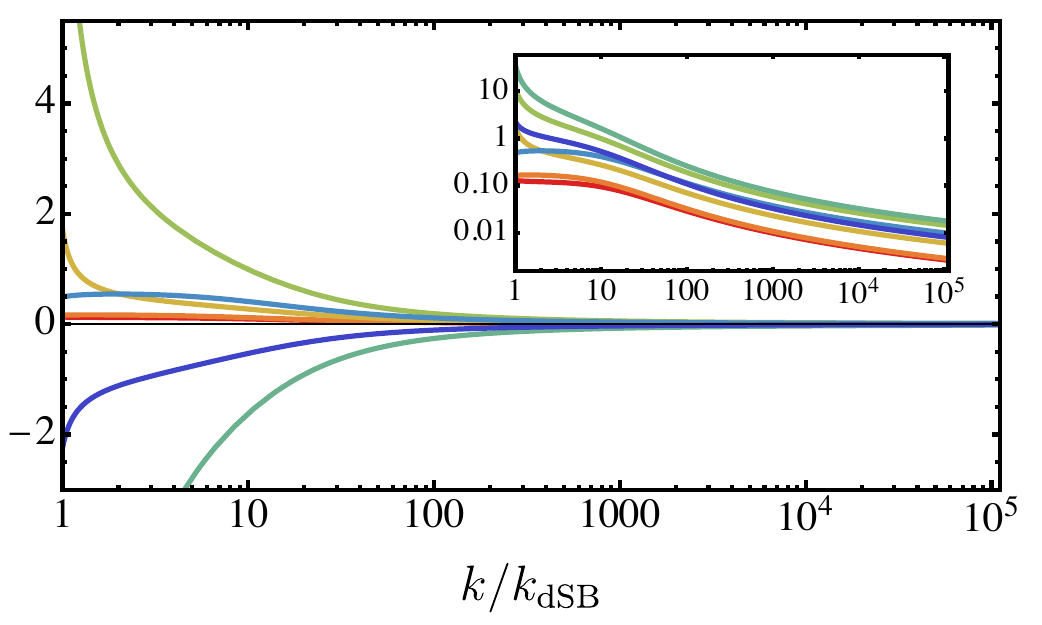}	
	\end{minipage}  
	\begin{minipage}{0.45\textwidth}
		\includegraphics[width=\columnwidth]{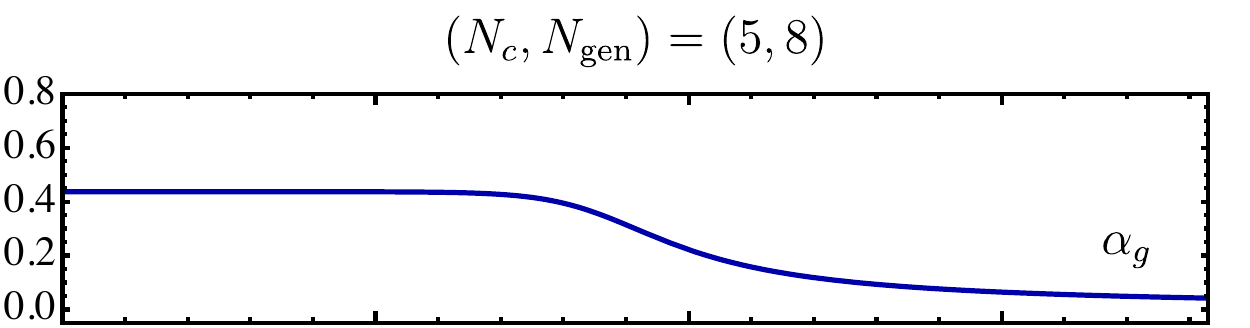}\vspace{-0.cm}
		\includegraphics[width=\columnwidth]{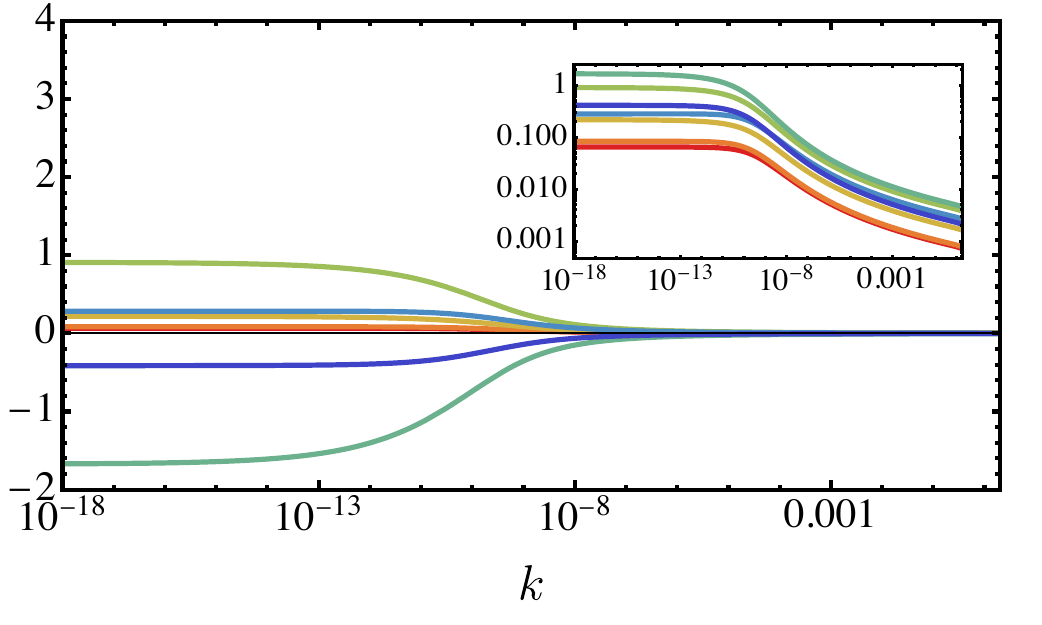}	
	\end{minipage}
	\hfil
	
	\caption{
		Integrated RG flows for different theories, specified by various values of $N_c$ and $\Ngen$. Top panels show the evolution of the gauge coupling; bottom panels show the four-fermion couplings defined in~\eqref{eq:Fierzbasis}. Insets display the absolute value of the four-fermion couplings on a logarithmic scale.}
	\label{fig: chiralSB explained}
\end{figure*}

Having set the stage, we now integrate the coupled system of four-fermion and gauge flows for different values of $N_c$ and $\Ngen$, as shown in \Cref{fig: chiralSB explained}. As detailed in the main text, the emergence of an IR fixed point in the flow of the four-fermion couplings indicates that these are fully determined by gauge dynamics. Furthermore, setting different UV boundary conditions does not affect the IR behavior: the system quickly approaches a partial fixed point, as observed in related contexts~\cite{Goertz:2024dnz,Braun:2014ata}. However, if the UV boundary conditions are too large, e.g., $\bar \lambda_{i,\,{k\to \infty}}\sim \mathcal{O}(10)$, the theory exits the universality class of perturbatively renormalizable gauge-fermion systems and enters the domain of nonperturbatively non-renormalizable gauge-NJL models. In that case, the IR behavior depends on the UV data, reflecting the effective nature of these theories. To avoid such effects, we take $\bar \lambda_{i,\,{k\to \infty}} = 0$ and $\alpha_{g,\,{k\to \infty}} \simeq 0$.

As the theory flows away from the UV Gaussian fixed point, an infinite tower of operators is dynamically generated, which is reflected in the growth of their associated couplings. The precise behavior depends on the symmetry structure and corresponding quantum fluctuations, as illustrated in \Cref{fig: chiralSB explained}.

\begin{figure}[t!]
	\centering
	\vspace{0.5cm}
	\includegraphics[width=0.5\columnwidth]{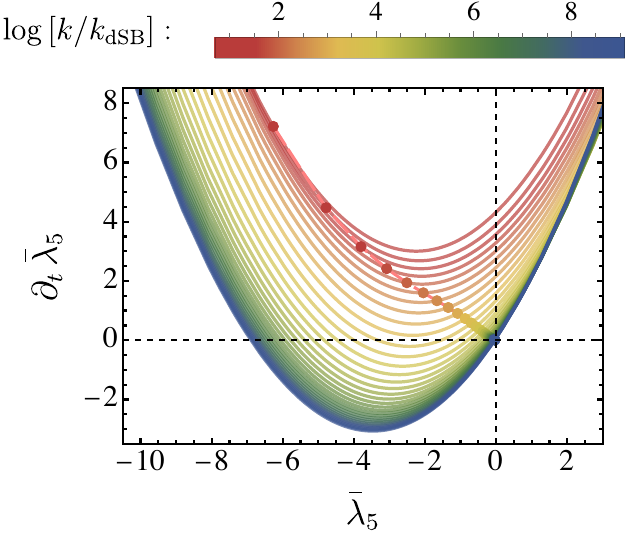}	
	\caption{Flow of the  $\bar \lambda_5$ dressing as a function of itself evaluated at different RG times for a theory with $N_c = 5$ and $\Ngen = 7$. Different curves and points indicate the running and position of the coupling at different scales.}
	\label{fig: flowlambda}
\end{figure}

\begin{figure*}[t!]
	\centering
	\includegraphics[width=.32\textwidth]{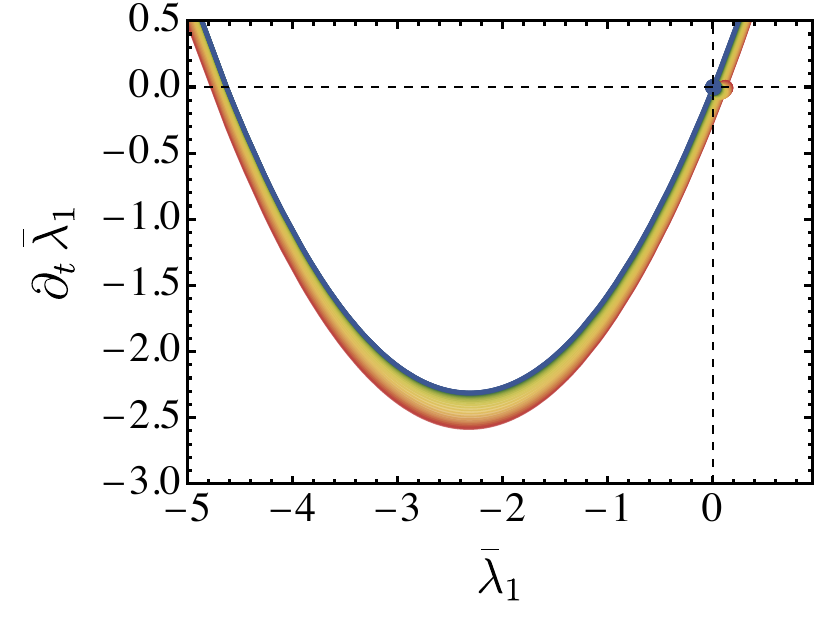}	
	\includegraphics[width=.32\textwidth]{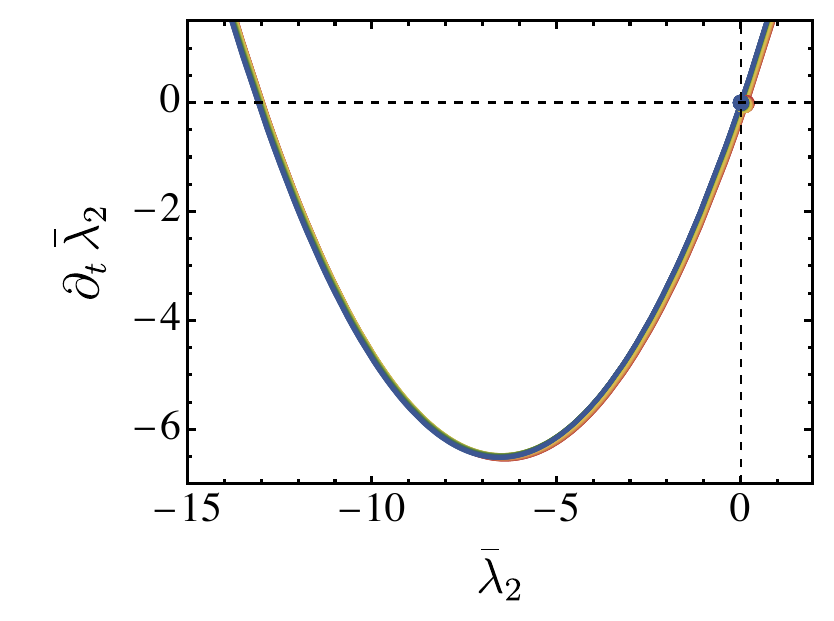}	
	\includegraphics[width=.32\textwidth]{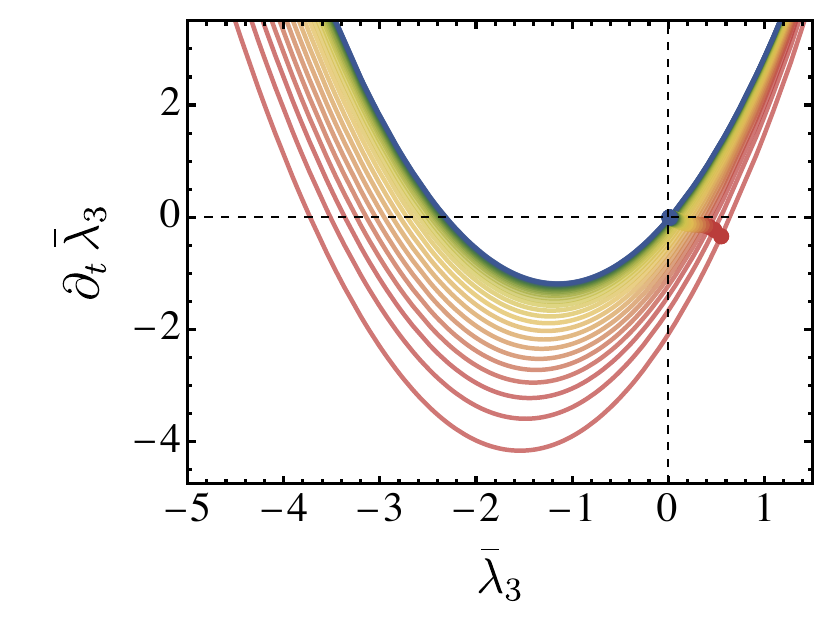}	
	\includegraphics[width=.32\textwidth]{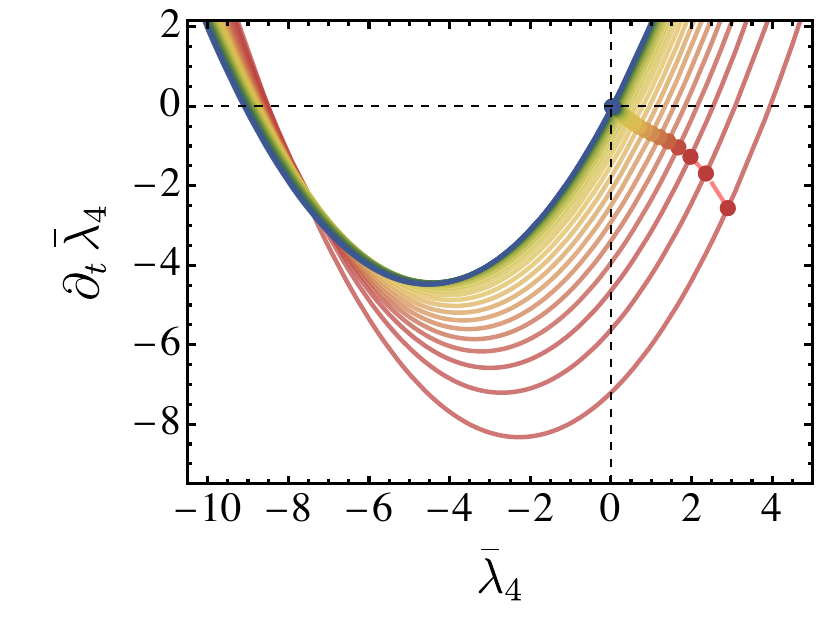}	
	\includegraphics[width=.32\textwidth]{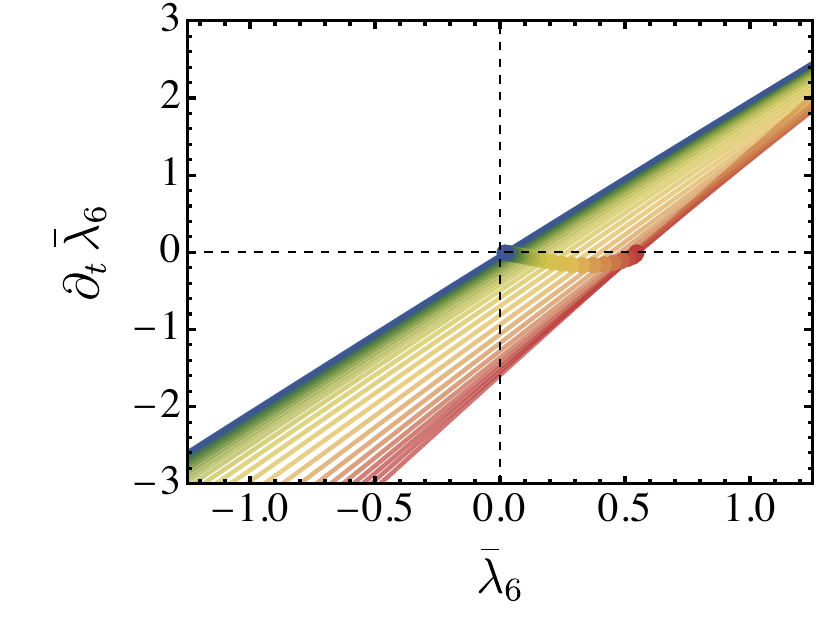}	
	\includegraphics[width=.32\textwidth]{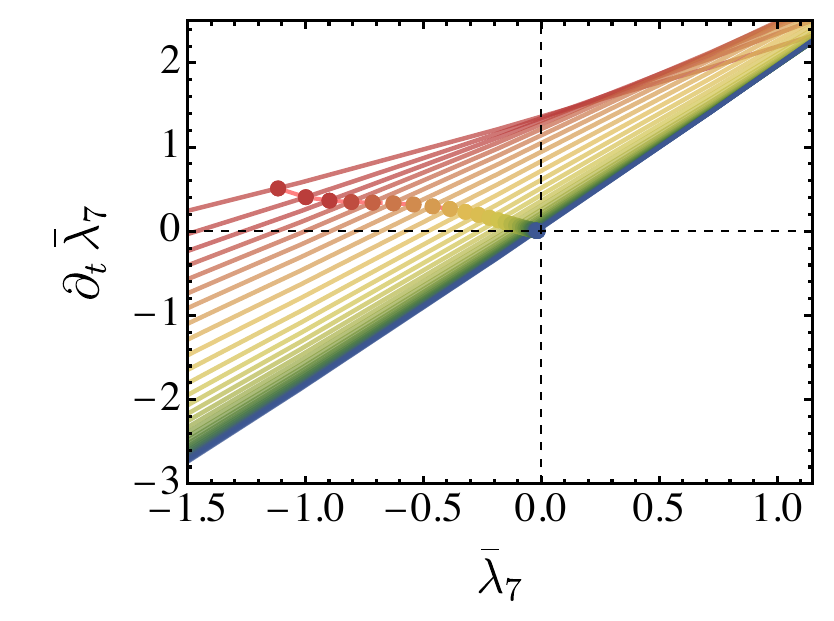}	
	\caption{Same as \Cref{fig: flowlambda}, but for the remaining four-fermion dressings defined in~\eqref{eq:Fierzbasis}.}
	\label{fig: flowlambdaall}
\end{figure*}

Due to the self-interaction terms $\boldsymbol{c}_{{\rm C },iii}$, the flow equations take a parabolic form (see \Cref{fig: flowlambda,fig: flowlambdaall}), where the direction of the parabola depends on the sign of the coefficient. Starting from the UV, the flow for small $\alpha_g$ always admits an IR attractive fixed point, which continuously connects to $\bar \lambda_{i,\,{k\to \infty}}= 0$ as $\alpha_g \to 0$. This behavior is insensitive to the coefficients $\boldsymbol{c}_{{\rm A}}, \boldsymbol{c}_{{\rm B}}, \boldsymbol{c}_{{\rm C}}$, provided $\alpha_g$ remains sufficiently small.

However, if any of the couplings displays a resonant structure (see \Cref{sec:phasediagram}) and $\alpha_g$ reaches $\acrit$, the corresponding IR fixed point disappears via a merger and a singularity appears in the RG flow. As shown in the main text, for $N_c = 5$, only two channels meet this condition: ${\cal O}_5$ and ${\cal O}_7$.

More precisely, $\acrit$ is reached first, and the singularity appears at a slightly lower scale, signaling the onset of dynamical symmetry breaking at the scale $\kSSB$, see~\cite{Goertz:2024dnz,Ihssen:2024miv}. Note that the scale where $\acrit$ is reached and $\kSSB$ are not identical. The gap between them can grow if other competing dynamics, such as confinement, become important, particularly when the system lies far from the conformal window.

Moving across theory space, the evolution of the gauge coupling changes, as fermionic contributions decelerate its IR growth. Eventually, as $\Ngen$ increases, the gauge coupling flows to an IR fixed point that remains below $\acrit$, and no dynamical scale is generated. This scenario is shown in the right-most panel of \Cref{fig: chiralSB explained}, where all four-fermion couplings approach scale-invariant values as $k \to 0$. Conversely, slightly decreasing $\Ngen$ may place the IR fixed point just above $\acrit$, leading to walking behavior and eventual condensation at $\kSSB$, as illustrated in the central panel of \Cref{fig: chiralSB explained} and discussed in~\cite{Goertz:2024dnz,Miransky1997,Miransky:1998dh,Braun:2006jd,Braun:2009ns,Braun:2010qs}. Furthermore, as $\kSSB$ is approached, all of the couplings start rapidly growing, see left and middle plots in \Cref{fig: chiralSB explained}. This is an artifact of the four-fermion computation. When bosonising, even partially only the dominant channel, it is known that the subleading tensor structures are not driven to a singularity, see e.g.~\cite{Goertz:2024dnz,Jaeckel:2002rm,Denz:2019ogb,Ihssen:2024miv}. 

To better visualize the dynamical symmetry breaking, we show in \Cref{fig: flowlambda,fig: flowlambdaall} the beta functions of the four-fermion couplings plotted against themselves for the case $N_c=5$, $\Ngen=7$ (central panel of \Cref{fig: chiralSB explained}). The flows are displayed at different RG times, from UV (blue) to IR (red), terminating at $k = \kSSB$. Here we see that the fixed-point merger, which as we discussed, is predominantly driven by the lowest-order gauge corrections and to a lesser extent by higher-order corrections via other channels.

By inspecting these flows, it is straightforward to identify the resonant channels. For example, in the cases of $\bar\lambda_5$ and $\bar\lambda_7$, the fixed point disappears in the IR, whereas for other couplings the fixed point persists. As a result, no condensation is expected in channels like ${\cal O}_1$ and ${\cal O}_2$, which only involve $\psi$ fields.


Finally, we comment on the signs of the four-fermion couplings defined in~\eqref{eq:effactionFermi}. These are dictated by the lowest-order contributions shown in \Cref{fig:cAcoefficients,eq:cAcoefficients}. For $N_c = 5$, all couplings except $\bar\lambda_5$ and $\bar\lambda_7$ have positive signs in our convention. This choice is motivated by bosonization arguments, where positive sign dressings correspond to a positive curvature of the mesonic potential in the chirally symmetric regime. These non-resonant channels reflect precisely this behavior. Conversely, the resonant channels exhibit negative sign dressings, which in a naive bosonization scheme would imply a tachyonic instability. However, this is not a concern here, as bosonization itself can introduce additional negative signs. Moreover, similar behavior is observed in QCD, where negative-sign couplings coexist with well-understood IR physics.

\section{Gauge beta functions and infrared fixed points}\label{app:betafunctionsandFP}
The $\overline{\mathrm{MS}}$ beta function for a $SU(N_c)$ chiral gauge theory can be written as
\begin{align}
	\label{eq:betafunctiong_perturbation}
	\beta(g)= &-g\,\sum_{n=1}^{\infty}  \left(\frac{g^2}{16 \pi^2 }\right)^{n}\,\beta_{n}\,,
\end{align}
with the first four-loop coefficients given by
\begin{subequations}
\begin{align}
\beta_0 &= \frac{11}{3}\,C_2(G) - \frac{2}{3}\sum_{i} T(R_{i}) N_{R_i}\,, \label{eq:beta0} \\[2ex]
\beta_1 &= \frac{34}{3}\,C_2(G)^2 - \sum_i \left[2\,C_2(R_i) + \frac{10}{3}\,C_2(G)\right]T(R_i) N_{R_i}\\[2ex]
\beta_2 &= \frac{2857}{54}\,C_2(G)^3  - \sum_i \left[-C_2(R_i)^2 + \frac{205}{18}\,C_2(G)C_2(R_i)+\frac{1415}{6}\,C_2(G)^3\right]T(R_i) N_{R_i} \notag\\[-.75ex]
&\;\;\;\;\; + \sum_{i,j} \left[ \frac{11}{9}\,C_2(R_i) + \frac{79}{54}\,C_2(G)\right]T(R_i) T(R_{j}) N_{R_i} N_{R_j} 
\end{align}

\begin{align}
\beta_3 &= \left( \frac{150653}{486}-\frac{44}{9}\zeta_3\right) C_2(G)^4  -\left( \frac{80}{9}-\frac{704}{3} \zeta_3 \right)\frac{d_{AA}}{{N_c}^2-1} \notag \\[-1ex]
&\;\;\;\;\; - \sum_i \left[-23 C_2(R_i)^3 + \left( \frac{4204}{54}-\frac{352}{18}\zeta_3\right)C_2(G)C_2(R_i)^2 \right. \notag \\[1ex]&\;\;\;\;\;- \left. \left( \frac{7073}{486}-\frac{656}{18}\zeta_3\right)C_2(G)^2 C_2(R_i) + \left( \frac{39143}{162}-\frac{136}{6} \zeta_3\right)\,C_2(G)^3\right]T(R_i) N_{R_i} \notag \\[1ex]& \;\;\;\;\;+ \sum_i N_{R_i} \left( \frac{512}{18}-\frac{1664}{6}\zeta_3\right) \frac{d_{R_i A}}{{N_c}^2-1}\notag\\[-.75ex]
&\;\;\;\;\; - \sum_{i,j} \left[ - \left(\frac{184}{12} -  \frac{64}{4}\zeta_3 \right) C_2(R_i) C_2(R_{j}) + \left(\frac{304}{108} +  \frac{128}{36}\zeta_3 \right) C_2(R_i)^2 \right. \notag \\[1ex] &\;\;\;\;\;-  \left. \left(\frac{17152}{972} +  \frac{448}{36}\zeta_3 \right) C_2(R_i) C_2(G) \right.-   \left.  \left( \frac{7930}{324} +  \frac{224}{36}\zeta_3 \right)  C_2(G)^2  \right]T(R_i) T(R_{j}) N_{R_i} N_{R_j}  \notag \\[1ex] &\;\;\;\;\; - \sum_{i,j} N_{R_i} N_{R_j} \left( \frac{704}{36} -\frac{512}{12}\zeta_3\right) \frac{d_{R_i R_{j}}}{{N_c}^2-1} \notag\\[-.75ex]
& \;\;\;\;\; +   \sum_{i,j,k} N_{R_i} N_{R_j} N_{R_k} T(R_{i})T(R_{j})T(R_{k}) \left(  \frac{154}{243}C_2(R_i) \frac{53}{243}C_2(G) \right)\,.\label{eq:beta1}
\end{align}
\end{subequations}
Here, the sum runs over all representations of the theory in terms of left-handed Weyl fermions. 
In our setup we have $\{ R_i \}=\{ \tiny \yng(1,1), \tiny  \bar{\yng(1)}\}$, with the respective multiplicities $N_{\tiny \yng(1,1)}= \Ngen$ and $N_{ \tiny \bar{\yng(1)}}= \Ngen(N_c-4)$.
Table~\ref{tab:groupdata1} summarizes the values of the dimension, the quadratic Casimir \(C_2(R)\) and the Dynkin index \(T(R)\) for relevant representations of \(\mathrm{SU}(N)\), and Table~\ref{tab:groupdata2} contains the coefficient $d_{R_i R_{j}}$  entering in the beta function.
\begin{table*}[h]
\centering
\begin{tabular}{|l|c|c|c|}
\hline
\rule{0pt}{4ex}    Representation & Dimension & Quadratic Casimir \(C_2(R)\) & Dynkin Index \(T(R)\)  \\[1ex] \hline
Anti-Fundamental  & \(N\) & \(\displaystyle \frac{N^2-1}{2N}\) & \(\displaystyle \frac{1}{2}\) \\[2ex]
Adjoint & \(N^2-1\) & \(N\) & \(N\) \\[2ex]
Antisymmetric  & \(\displaystyle \frac{N(N-1)}{2}\) & \(\displaystyle \frac{(N-2)(N+1)}{N}\) & \(\displaystyle \frac{N-2}{2}\) \\[2ex]
Symmetric  & \(\displaystyle \frac{N(N+1)}{2}\) & \(\displaystyle \frac{(N+2)(N-1)}{N}\) & \(\displaystyle \frac{N+2}{2}\) \\[2ex]
\hline
\end{tabular}
\caption{Dimensions, quadratic Casimirs and Dynkin indices for various representations of \(\mathrm{SU}(N)\). Conjugate representations have identical values.}
\label{tab:groupdata1}
\end{table*}

\begin{table*}[h]
\centering
\begin{adjustbox}{angle=-90}
\begin{tabular}{|c|c|c|c|}
\hline  \rule{0pt}{4ex}  & \rule{5pt}{0ex} Anti-Fund \rule{5pt}{0ex}& Adjoint & Antisymmetric \\[2ex] \hline
\rule{0pt}{4ex} Anti-Fund  & \( \displaystyle \frac{({N }^6 - 7 {N }^4 + 24 {N }^2 - 18)}{96 {N }^2}\) & \( \displaystyle \frac{{N }({N }^4 +5 {N }^2 -6)}{48}\) &  \( \displaystyle \frac{({N }-2)({N }^2-1)({N }(N -6)^2(N +6) +72)}{96 {N }^2}\) \\[2ex]
\rule{0pt}{2ex} Adjoint &  & \( \displaystyle \frac{{N }^2({N }^4 +35 {N }^2 -36)}{24}\) &\( \displaystyle \frac{({N }-2)({N }^2-1){N }((N -6)N +24)}{48}\)  \\[2ex]
\rule{3pt}{0ex} Antisymmetric \rule{0pt}{0ex}  & & & $\frac{(N-2) (N-1) (N+1) ((N-2) N (N ((N-12) N+48)+144)-576)}{96 N^2}$ \\[2ex]
\hline
\end{tabular}
\end{adjustbox}
\caption{Group coefficient $d_{R_f R_{f'}}$ for different \(\mathrm{SU}(N)\) representations.}
\label{tab:groupdata2}
\end{table*}
\bibliographystyle{JHEP}
\bibliography{fRG_chiral}

\providecommand{\href}[2]{#2}\begingroup\raggedright\begin{thebibliography}{10}

\bibitem{tHooft:1979rat}
G.~'t~Hooft, \emph{{Naturalness, chiral symmetry, and spontaneous chiral symmetry breaking}}, \href{https://doi.org/10.1007/978-1-4684-7571-5\_9}{\emph{NATO Sci. Ser. B} {\bfseries 59} (1980) 135}.

\bibitem{Seiberg:1994rs}
N.~Seiberg and E.~Witten, \emph{{Electric - magnetic duality, monopole condensation, and confinement in N=2 supersymmetric Yang-Mills theory}}, \href{https://doi.org/10.1016/0550-3213(94)90124-4}{\emph{Nucl. Phys. B} {\bfseries 426} (1994) 19} [\href{https://arxiv.org/abs/hep-th/9407087}{{\ttfamily hep-th/9407087}}].

\bibitem{Seiberg:1994aj}
N.~Seiberg and E.~Witten, \emph{{Monopoles, duality and chiral symmetry breaking in N=2 supersymmetric QCD}}, \href{https://doi.org/10.1016/0550-3213(94)90214-3}{\emph{Nucl. Phys. B} {\bfseries 431} (1994) 484} [\href{https://arxiv.org/abs/hep-th/9408099}{{\ttfamily hep-th/9408099}}].

\bibitem{Seiberg:1994bz}
N.~Seiberg, \emph{{Exact results on the space of vacua of four-dimensional SUSY gauge theories}}, \href{https://doi.org/10.1103/PhysRevD.49.6857}{\emph{Phys. Rev. D} {\bfseries 49} (1994) 6857} [\href{https://arxiv.org/abs/hep-th/9402044}{{\ttfamily hep-th/9402044}}].

\bibitem{Seiberg:1994pq}
N.~Seiberg, \emph{{Electric - magnetic duality in supersymmetric nonAbelian gauge theories}}, \href{https://doi.org/10.1016/0550-3213(94)00023-8}{\emph{Nucl. Phys. B} {\bfseries 435} (1995) 129} [\href{https://arxiv.org/abs/hep-th/9411149}{{\ttfamily hep-th/9411149}}].

\bibitem{Ciambriello:2022wmh}
L.~Ciambriello, R.~Contino, A.~Luzio, M.~Romano and L.-X.~Xu, \emph{{Proof of chiral symmetry breaking from anomaly matching in QCD-like theories}}, \href{https://doi.org/10.1103/PhysRevD.110.114035}{\emph{Phys. Rev. D} {\bfseries 110} (2024) 114035} [\href{https://arxiv.org/abs/2212.02930}{{\ttfamily 2212.02930}}].

\bibitem{Ciambriello:2024xzd}
L.~Ciambriello, R.~Contino, A.~Luzio, M.~Romano and L.-X.~Xu, \emph{{A novel strategy to prove chiral symmetry breaking in QCD-like theories}}, \href{https://doi.org/10.1016/j.physletb.2025.139367}{\emph{Phys. Lett. B} {\bfseries 862} (2025) 139367} [\href{https://arxiv.org/abs/2404.02967}{{\ttfamily 2404.02967}}].

\bibitem{Vafa:1983tf}
C.~Vafa and E.~Witten, \emph{{Restrictions on Symmetry Breaking in Vector-Like Gauge Theories}}, \href{https://doi.org/10.1016/0550-3213(84)90230-X}{\emph{Nucl. Phys. B} {\bfseries 234} (1984) 173}.

\bibitem{Wilson:1974sk}
K.G.~Wilson, \emph{{Confinement of Quarks}}, \href{https://doi.org/10.1103/PhysRevD.10.2445}{\emph{Phys. Rev. D} {\bfseries 10} (1974) 2445}.

\bibitem{Kogut:1982ds}
J.B.~Kogut, \emph{{A Review of the Lattice Gauge Theory Approach to Quantum Chromodynamics}}, \href{https://doi.org/10.1103/RevModPhys.55.775}{\emph{Rev. Mod. Phys.} {\bfseries 55} (1983) 775}.

\bibitem{Greensite:2003bk}
J.~Greensite, \emph{{The confinement problem in lattice gauge theory}}, \href{https://doi.org/10.1016/S0146-6410(03)90012-3}{\emph{Prog. Part. Nucl. Phys.} {\bfseries 51} (2003) 1} [\href{https://arxiv.org/abs/hep-lat/0301023}{{\ttfamily hep-lat/0301023}}].

\bibitem{Bogolubsky:2009dc}
I.L.~Bogolubsky, E.M.~Ilgenfritz, M.~M{\"u}ller-Preussker and A.~Sternbeck, \emph{{Lattice gluodynamics computation of Landau gauge Green's functions in the deep infrared}}, \href{https://doi.org/10.1016/j.physletb.2009.04.076}{\emph{Phys. Lett.} {\bfseries B676} (2009) 69} [\href{https://arxiv.org/abs/0901.0736}{{\ttfamily 0901.0736}}].

\bibitem{Bowman:2004jm}
P.O.~Bowman, U.M.~Heller, D.B.~Leinweber, M.B.~Parappilly and A.G.~Williams, \emph{{Unquenched gluon propagator in Landau gauge}}, \href{https://doi.org/10.1103/PhysRevD.70.034509}{\emph{Phys. Rev. D} {\bfseries 70} (2004) 034509} [\href{https://arxiv.org/abs/hep-lat/0402032}{{\ttfamily hep-lat/0402032}}].

\bibitem{Aoki:2009sc}
Y.~Aoki, S.~Borsanyi, S.~Durr, Z.~Fodor, S.D.~Katz et~al., \emph{{The QCD transition temperature: results with physical masses in the continuum limit II.}}, \href{https://doi.org/10.1088/1126-6708/2009/06/088}{\emph{JHEP} {\bfseries 0906} (2009) 088} [\href{https://arxiv.org/abs/0903.4155}{{\ttfamily 0903.4155}}].

\bibitem{Alkofer:2000wg}
R.~Alkofer and L.~von Smekal, \emph{{The infrared behavior of QCD Green's functions: Confinement, dynamical symmetry breaking, and hadrons as relativistic bound states}}, \href{https://doi.org/10.1016/S0370-1573(01)00010-2}{\emph{Phys. Rept.} {\bfseries 353} (2001) 281} [\href{https://arxiv.org/abs/hep-ph/0007355}{{\ttfamily hep-ph/0007355}}].

\bibitem{Fischer:2006ub}
C.S.~Fischer, \emph{{Infrared properties of QCD from Dyson-Schwinger equations}}, \href{https://doi.org/10.1088/0954-3899/32/8/R02}{\emph{J.Phys.G} {\bfseries G32} (2006) R253} [\href{https://arxiv.org/abs/hep-ph/0605173}{{\ttfamily hep-ph/0605173}}].

\bibitem{Pawlowski:2003hq}
J.M.~Pawlowski, D.F.~Litim, S.~Nedelko and L.~von Smekal, \emph{{Infrared behavior and fixed points in Landau gauge QCD}}, \href{https://doi.org/10.1103/PhysRevLett.93.152002}{\emph{Phys.Rev.Lett.} {\bfseries 93} (2004) 152002} [\href{https://arxiv.org/abs/hep-th/0312324}{{\ttfamily hep-th/0312324}}].

\bibitem{Braun:2007bx}
J.~Braun, H.~Gies and J.M.~Pawlowski, \emph{{Quark Confinement from Color Confinement}}, \href{https://doi.org/10.1016/j.physletb.2010.01.009}{\emph{Phys.Lett.} {\bfseries B684} (2010) 262} [\href{https://arxiv.org/abs/0708.2413}{{\ttfamily 0708.2413}}].

\bibitem{Cyrol:2016tym}
A.K.~Cyrol, L.~Fister, M.~Mitter, J.M.~Pawlowski and N.~Strodthoff, \emph{{Landau gauge Yang-Mills correlation functions}}, \href{https://doi.org/10.1103/PhysRevD.94.054005}{\emph{Phys. Rev.} {\bfseries D94} (2016) 054005} [\href{https://arxiv.org/abs/1605.01856}{{\ttfamily 1605.01856}}].

\bibitem{Aguilar:2022thg}
A.C.~Aguilar, F.~De~Soto, M.N.~Ferreira, J.~Papavassiliou, F.~Pinto-G\'omez, C.D.~Roberts et~al., \emph{{Schwinger mechanism for gluons from lattice QCD}}, \href{https://doi.org/10.1016/j.physletb.2023.137906}{\emph{Phys. Lett. B} {\bfseries 841} (2023) 137906} [\href{https://arxiv.org/abs/2211.12594}{{\ttfamily 2211.12594}}].

\bibitem{Gross:2022hyw}
F.~Gross et~al., \emph{{50 Years of Quantum Chromodynamics}}, \href{https://doi.org/10.1140/epjc/s10052-023-11949-2}{\emph{Eur. Phys. J. C} {\bfseries 83} (2023) 1125} [\href{https://arxiv.org/abs/2212.11107}{{\ttfamily 2212.11107}}].

\bibitem{Raby:1979my}
S.~Raby, S.~Dimopoulos and L.~Susskind, \emph{{Tumbling Gauge Theories}}, \href{https://doi.org/10.1016/0550-3213(80)90093-0}{\emph{Nucl. Phys. B} {\bfseries 169} (1980) 373}.

\bibitem{Dimopoulos:1980hn}
S.~Dimopoulos, S.~Raby and L.~Susskind, \emph{{Light Composite Fermions}}, \href{https://doi.org/10.1016/0550-3213(80)90215-1}{\emph{Nucl. Phys. B} {\bfseries 173} (1980) 208}.

\bibitem{Bars:1981se}
I.~Bars and S.~Yankielowicz, \emph{{Composite Quarks and Leptons as Solutions of Anomaly Constraints}}, \href{https://doi.org/10.1016/0370-2693(81)90664-X}{\emph{Phys. Lett. B} {\bfseries 101} (1981) 159}.

\bibitem{Nielsen:1980rz}
H.B.~Nielsen and M.~Ninomiya, \emph{{Absence of Neutrinos on a Lattice. 1. Proof by Homotopy Theory}}, \href{https://doi.org/10.1016/0550-3213(82)90011-6}{\emph{Nucl. Phys. B} {\bfseries 185} (1981) 20}.

\bibitem{Nielsen:1981hk}
H.B.~Nielsen and M.~Ninomiya, \emph{{No Go Theorem for Regularizing Chiral Fermions}}, \href{https://doi.org/10.1016/0370-2693(81)91026-1}{\emph{Phys. Lett. B} {\bfseries 105} (1981) 219}.

\bibitem{Nielsen:1981xu}
H.B.~Nielsen and M.~Ninomiya, \emph{{Absence of Neutrinos on a Lattice. 2. Intuitive Topological Proof}}, \href{https://doi.org/10.1016/0550-3213(81)90524-1}{\emph{Nucl. Phys. B} {\bfseries 193} (1981) 173}.

\bibitem{Bolognesi:2021jzs}
S.~Bolognesi, K.~Konishi and A.~Luzio, \emph{{Anomalies and phases of strongly coupled chiral gauge theories: Recent developments}}, \href{https://doi.org/10.1142/S0217751X22300149}{\emph{Int. J. Mod. Phys. A} {\bfseries 37} (2022) 2230014} [\href{https://arxiv.org/abs/2110.02104}{{\ttfamily 2110.02104}}].

\bibitem{Bolognesi:2023sxe}
S.~Bolognesi, K.~Konishi and A.~Luzio, \emph{{Dynamics of strongly-coupled chiral gauge theories}}, \href{https://doi.org/10.1088/1742-6596/2531/1/012006}{\emph{J. Phys. Conf. Ser.} {\bfseries 2531} (2023) 012006} [\href{https://arxiv.org/abs/2304.03357}{{\ttfamily 2304.03357}}].

\bibitem{Konishi:2024rjz}
K.~Konishi, S.~Bolognesi and A.~Luzio, \emph{{Anomalies and Dynamics in Strongly-Coupled Gauge Theories, New Criteria for Different Phases, and a Lesson from Supersymmetric Gauge Theories}}, \href{https://doi.org/10.22323/1.463.0126}{\emph{PoS} {\bfseries CORFU2023} (2024) 126} [\href{https://arxiv.org/abs/2403.15775}{{\ttfamily 2403.15775}}].

\bibitem{Sannino:2009za}
F.~Sannino, \emph{{Conformal Dynamics for TeV Physics and Cosmology}}, {\emph{Acta Phys. Polon. B} {\bfseries 40} (2009) 3533} [\href{https://arxiv.org/abs/0911.0931}{{\ttfamily 0911.0931}}].

\bibitem{Poppitz:2009uq}
E.~Poppitz and M.~Unsal, \emph{{Conformality or confinement: (IR)relevance of topological excitations}}, \href{https://doi.org/10.1088/1126-6708/2009/09/050}{\emph{JHEP} {\bfseries 09} (2009) 050} [\href{https://arxiv.org/abs/0906.5156}{{\ttfamily 0906.5156}}].

\bibitem{Bolognesi:2019wfq}
S.~Bolognesi and K.~Konishi, \emph{{Dynamics and symmetries in chiral $SU(N)$ gauge theories}}, \href{https://doi.org/10.1103/PhysRevD.100.114008}{\emph{Phys. Rev. D} {\bfseries 100} (2019) 114008} [\href{https://arxiv.org/abs/1906.01485}{{\ttfamily 1906.01485}}].

\bibitem{Bolognesi:2020mpe}
S.~Bolognesi, K.~Konishi and A.~Luzio, \emph{{Dynamics from symmetries in chiral $SU(N)$ gauge theories}}, \href{https://doi.org/10.1007/JHEP09(2020)001}{\emph{JHEP} {\bfseries 09} (2020) 001} [\href{https://arxiv.org/abs/2004.06639}{{\ttfamily 2004.06639}}].

\bibitem{Bolognesi:2021yni}
S.~Bolognesi, K.~Konishi and A.~Luzio, \emph{{Probing the dynamics of chiral $SU(N)$ gauge theories via generalized anomalies}}, \href{https://doi.org/10.1103/PhysRevD.103.094016}{\emph{Phys. Rev. D} {\bfseries 103} (2021) 094016} [\href{https://arxiv.org/abs/2101.02601}{{\ttfamily 2101.02601}}].

\bibitem{Bolognesi:2024bnm}
S.~Bolognesi, K.~Konishi, A.~Luzio and M.~Orso, \emph{{Natural Anomaly Matching}},  \href{https://arxiv.org/abs/2410.01315}{{\ttfamily 2410.01315}}.

\bibitem{Fradkin:1978dv}
E.H.~Fradkin and S.H.~Shenker, \emph{{Phase Diagrams of Lattice Gauge Theories with Higgs Fields}}, \href{https://doi.org/10.1103/PhysRevD.19.3682}{\emph{Phys. Rev.} {\bfseries D19} (1979) 3682}.

\bibitem{Wetterich:1992yh}
C.~Wetterich, \emph{{Exact evolution equation for the effective potential}}, \href{https://doi.org/10.1016/0370-2693(93)90726-X}{\emph{Phys. Lett.} {\bfseries B301} (1993) 90} [\href{https://arxiv.org/abs/1710.05815}{{\ttfamily 1710.05815}}].

\bibitem{Dupuis:2020fhh}
N.~Dupuis, L.~Canet, A.~Eichhorn, W.~Metzner, J.M.~Pawlowski, M.~Tissier et~al., \emph{{The nonperturbative functional renormalization group and its applications}}, \href{https://doi.org/10.1016/j.physrep.2021.01.001}{\emph{Phys. Rept.} {\bfseries 910} (2021) 1} [\href{https://arxiv.org/abs/2006.04853}{{\ttfamily 2006.04853}}].

\bibitem{Miransky:1998dh}
V.~Miransky, \emph{{Dynamics in the conformal window in QCD like theories}}, \href{https://doi.org/10.1103/PhysRevD.59.105003}{\emph{Phys.Rev.} {\bfseries D59} (1999) 105003} [\href{https://arxiv.org/abs/hep-ph/9812350}{{\ttfamily hep-ph/9812350}}].

\bibitem{Miransky1997}
V.A.~Miransky and K.~Yamawaki, \emph{{Conformal phase transition in gauge theories}}, \href{https://doi.org/10.1103/PhysRevD.56.3768}{\emph{Phys. Rev. D} {\bfseries 55} (1997) 5051} [\href{https://arxiv.org/abs/hep-th/9611142}{{\ttfamily hep-th/9611142}}].

\bibitem{Gies:2005as}
H.~Gies and J.~Jaeckel, \emph{{Chiral phase structure of QCD with many flavors}}, \href{https://doi.org/10.1140/epjc/s2006-02475-0}{\emph{Eur. Phys. J. C} {\bfseries 46} (2006) 433} [\href{https://arxiv.org/abs/hep-ph/0507171}{{\ttfamily hep-ph/0507171}}].

\bibitem{Braun:2006jd}
J.~Braun and H.~Gies, \emph{{Chiral phase boundary of QCD at finite temperature}}, \href{https://doi.org/10.1088/1126-6708/2006/06/024}{\emph{JHEP} {\bfseries 0606} (2006) 024} [\href{https://arxiv.org/abs/hep-ph/0602226}{{\ttfamily hep-ph/0602226}}].

\bibitem{Braun:2009ns}
J.~Braun and H.~Gies, \emph{{Scaling laws near the conformal window of many-flavor QCD}}, \href{https://doi.org/10.1007/JHEP05(2010)060}{\emph{JHEP} {\bfseries 05} (2010) 060} [\href{https://arxiv.org/abs/0912.4168}{{\ttfamily 0912.4168}}].

\bibitem{Goertz:2024dnz}
F.~Goertz, {\'A}.~Pastor-Guti{\'e}rrez and J.M.~Pawlowski, \emph{{Gauge-fermion cartography: From confinement and chiral symmetry breaking to conformality}}, \href{https://doi.org/10.1103/7dzj-k6k8}{\emph{Phys. Rev. D} {\bfseries 112} (2025) 034029} [\href{https://arxiv.org/abs/2412.12254}{{\ttfamily 2412.12254}}].

\bibitem{Braun:2010qs}
J.~Braun, C.S.~Fischer and H.~Gies, \emph{{Beyond Miransky Scaling}}, \href{https://doi.org/10.1103/PhysRevD.84.034045}{\emph{Phys. Rev. D} {\bfseries 84} (2011) 034045} [\href{https://arxiv.org/abs/1012.4279}{{\ttfamily 1012.4279}}].

\bibitem{Cyrol:2017ewj}
A.K.~Cyrol, M.~Mitter, J.M.~Pawlowski and N.~Strodthoff, \emph{{Nonperturbative quark, gluon, and meson correlators of unquenched QCD}}, \href{https://doi.org/10.1103/PhysRevD.97.054006}{\emph{Phys. Rev.} {\bfseries D97} (2018) 054006} [\href{https://arxiv.org/abs/1706.06326}{{\ttfamily 1706.06326}}].

\bibitem{Mitter:2014wpa}
M.~Mitter, J.M.~Pawlowski and N.~Strodthoff, \emph{{Chiral symmetry breaking in continuum QCD}}, \href{https://doi.org/10.1103/PhysRevD.91.054035}{\emph{Phys. Rev.} {\bfseries D91} (2015) 054035} [\href{https://arxiv.org/abs/1411.7978}{{\ttfamily 1411.7978}}].

\bibitem{Ihssen:2024miv}
F.~Ihssen, J.M.~Pawlowski, F.R.~Sattler and N.~Wink, \emph{Towards quantitative precision in functional qcd i},  \href{https://arxiv.org/abs/2408.08413}{{\ttfamily 2408.08413}}.

\bibitem{Fu:2019hdw}
W.-j.~Fu, J.M.~Pawlowski and F.~Rennecke, \emph{Qcd phase structure at finite temperature and density}, \href{https://doi.org/10.1103/PhysRevD.101.054032}{\emph{Phys. Rev. D} {\bfseries 101} (2020) 054032} [\href{https://arxiv.org/abs/1909.02991}{{\ttfamily 1909.02991}}].

\bibitem{Georgi:1974sy}
H.~Georgi and S.L.~Glashow, \emph{{Unity of All Elementary Particle Forces}}, \href{https://doi.org/10.1103/PhysRevLett.32.438}{\emph{Phys. Rev. Lett.} {\bfseries 32} (1974) 438}.

\bibitem{Caswell:1974gg}
W.E.~Caswell, \emph{{Asymptotic Behavior of Nonabelian Gauge Theories to Two Loop Order}}, \href{https://doi.org/10.1103/PhysRevLett.33.244}{\emph{Phys. Rev. Lett.} {\bfseries 33} (1974) 244}.

\bibitem{Banks:1981nn}
T.~Banks and A.~Zaks, \emph{{On the Phase Structure of Vector-Like Gauge Theories with Massless Fermions}}, \href{https://doi.org/10.1016/0550-3213(82)90035-9}{\emph{Nucl. Phys. B} {\bfseries 196} (1982) 189}.

\bibitem{Bai:2021tgl}
Y.~Bai and D.~Stolarski, \emph{{Phases of confining SU(5) chiral gauge theory with three generations}}, \href{https://doi.org/10.1007/JHEP03(2022)113}{\emph{JHEP} {\bfseries 03} (2022) 113} [\href{https://arxiv.org/abs/2111.11214}{{\ttfamily 2111.11214}}].

\bibitem{Nambu:1961fr}
Y.~Nambu and G.~Jona-Lasinio, \emph{{Dynamical model of elementary particles based on an analogy with superconductivity. II.}}, \href{https://doi.org/10.1103/PhysRev.124.246}{\emph{Phys. Rev.} {\bfseries 124} (1961) 246}.

\bibitem{Braun:2011pp}
J.~Braun, \emph{{Fermion Interactions and Universal Behavior in Strongly Interacting Theories}}, \href{https://doi.org/10.1088/0954-3899/39/3/033001}{\emph{J.Phys.} {\bfseries G39} (2012) 033001} [\href{https://arxiv.org/abs/1108.4449}{{\ttfamily 1108.4449}}].

\bibitem{HubbardPhysRevLett.3.77}
J.~Hubbard, \emph{Calculation of partition functions}, \href{https://doi.org/10.1103/PhysRevLett.3.77}{\emph{Phys. Rev. Lett.} {\bfseries 3} (1959) 77}.

\bibitem{Stratonovich}
R.L.~{Stratonovich}, \emph{{On a Method of Calculating Quantum Distribution Functions}}, {\emph{Soviet Physics Doklady} {\bfseries 2} (1957) 416}.

\bibitem{Miransky1989b}
V.A.~Miransky and K.~Yamawaki, \emph{{On Gauge Theories with Additional Four Fermion Interaction}}, \href{https://doi.org/10.1142/S0217732389000186}{\emph{Mod. Phys. Lett. A} {\bfseries 4} (1989) 129}.

\bibitem{Miransky:1994vk}
V.A.~Miransky, \emph{{Dynamical symmetry breaking in quantum field theories}} (1994).

\bibitem{Wetterich:1989xg}
C.~Wetterich, \emph{{Average Action and the Renormalization Group Equations}}, \href{https://doi.org/10.1016/0550-3213(91)90099-J}{\emph{Nucl. Phys. B} {\bfseries 352} (1991) 529}.

\bibitem{Wetterich:1991be}
C.~Wetterich, \emph{{The Average action for scalar fields near phase transitions}}, \href{https://doi.org/10.1007/BF01474340}{\emph{Z.Phys.} {\bfseries C57} (1993) 451}.

\bibitem{Litim:2001up}
D.F.~Litim, \emph{{Optimized renormalization group flows}}, \href{https://doi.org/10.1103/PhysRevD.64.105007}{\emph{Phys.Rev.} {\bfseries D64} (2001) 105007} [\href{https://arxiv.org/abs/hep-th/0103195}{{\ttfamily hep-th/0103195}}].

\bibitem{Fu:2022uow}
W.-j.~Fu, C.~Huang, J.M.~Pawlowski and Y.-y.~Tan, \emph{{Four-quark scatterings in QCD I}}, \href{https://doi.org/10.21468/SciPostPhys.14.4.069}{\emph{SciPost Phys.} {\bfseries 14} (2023) 069} [\href{https://arxiv.org/abs/2209.13120}{{\ttfamily 2209.13120}}].

\bibitem{Fu:2024ysj}
W.-j.~Fu, C.~Huang, J.M.~Pawlowski and Y.-y.~Tan, \emph{{Four-quark scatterings in QCD II}},  \href{https://arxiv.org/abs/2401.07638}{{\ttfamily 2401.07638}}.

\bibitem{Fu:2025hcm}
W.-j.~Fu, C.~Huang, J.M.~Pawlowski, Y.-y.~Tan and L.-j.~Zhou, \emph{{Four-quark scatterings in QCD III}},  \href{https://arxiv.org/abs/2502.14388}{{\ttfamily 2502.14388}}.

\bibitem{Braun:2018bik}
J.~Braun, M.~Leonhardt and M.~Pospiech, \emph{{Fierz-complete NJL model study. II. Toward the fixed-point and phase structure of hot and dense two-flavor QCD}}, \href{https://doi.org/10.1103/PhysRevD.97.076010}{\emph{Phys. Rev. D} {\bfseries 97} (2018) 076010} [\href{https://arxiv.org/abs/1801.08338}{{\ttfamily 1801.08338}}].

\bibitem{Braun:2019aow}
J.~Braun, M.~Leonhardt and M.~Pospiech, \emph{{Fierz-complete NJL model study III: Emergence from quark-gluon dynamics}}, \href{https://doi.org/10.1103/PhysRevD.101.036004}{\emph{Phys. Rev. D} {\bfseries 101} (2020) 036004} [\href{https://arxiv.org/abs/1909.06298}{{\ttfamily 1909.06298}}].

\bibitem{Gies:2002hq}
H.~Gies and C.~Wetterich, \emph{{Universality of spontaneous chiral symmetry breaking in gauge theories}}, \href{https://doi.org/10.1103/PhysRevD.69.025001}{\emph{Phys.Rev.} {\bfseries D69} (2004) 025001} [\href{https://arxiv.org/abs/hep-th/0209183}{{\ttfamily hep-th/0209183}}].

\bibitem{Fukushima:2021ctq}
K.~Fukushima, J.M.~Pawlowski and N.~Strodthoff, \emph{{Emergent Hadrons and Diquarks}},  \href{https://arxiv.org/abs/2103.01129}{{\ttfamily 2103.01129}}.

\bibitem{Pawlowski:2005xe}
J.M.~Pawlowski, \emph{{Aspects of the functional renormalisation group}}, \href{https://doi.org/10.1016/j.aop.2007.01.007}{\emph{Annals Phys.} {\bfseries 322} (2007) 2831} [\href{https://arxiv.org/abs/hep-th/0512261}{{\ttfamily hep-th/0512261}}].

\bibitem{Braun:2005uj}
J.~Braun and H.~Gies, \emph{{Running coupling at finite temperature and chiral symmetry restoration in QCD}}, \href{https://doi.org/10.1016/j.physletb.2006.11.059}{\emph{Phys. Lett.} {\bfseries B645} (2007) 53} [\href{https://arxiv.org/abs/hep-ph/0512085}{{\ttfamily hep-ph/0512085}}].

\bibitem{Li:2020gnx}
H.-L.~Li, Z.~Ren, J.~Shu, M.-L.~Xiao, J.-H.~Yu and Y.-H.~Zheng, \emph{{Complete set of dimension-eight operators in the standard model effective field theory}}, \href{https://doi.org/10.1103/PhysRevD.104.015026}{\emph{Phys. Rev. D} {\bfseries 104} (2021) 015026} [\href{https://arxiv.org/abs/2005.00008}{{\ttfamily 2005.00008}}].

\bibitem{Li:2020xlh}
H.-L.~Li, Z.~Ren, M.-L.~Xiao, J.-H.~Yu and Y.-H.~Zheng, \emph{{Complete set of dimension-nine operators in the standard model effective field theory}}, \href{https://doi.org/10.1103/PhysRevD.104.015025}{\emph{Phys. Rev. D} {\bfseries 104} (2021) 015025} [\href{https://arxiv.org/abs/2007.07899}{{\ttfamily 2007.07899}}].

\bibitem{Li:2022tec}
H.-L.~Li, Z.~Ren, M.-L.~Xiao, J.-H.~Yu and Y.-H.~Zheng, \emph{{Operators for generic effective field theory at any dimension: on-shell amplitude basis construction}}, \href{https://doi.org/10.1007/JHEP04(2022)140}{\emph{JHEP} {\bfseries 04} (2022) 140} [\href{https://arxiv.org/abs/2201.04639}{{\ttfamily 2201.04639}}].

\bibitem{Ferreira:2025anh}
M.N.~Ferreira and J.~Papavassiliou, \emph{Gluon mass scale through the schwinger mechanism},  \href{https://arxiv.org/abs/2501.01080}{{\ttfamily 2501.01080}}.

\bibitem{Fukushima:2010bq}
K.~Fukushima and T.~Hatsuda, \emph{{The phase diagram of dense QCD}}, \href{https://doi.org/10.1088/0034-4885/74/1/014001}{\emph{Rept.Prog.Phys.} {\bfseries 74} (2011) 014001} [\href{https://arxiv.org/abs/1005.4814}{{\ttfamily 1005.4814}}].

\bibitem{Alford:2007xm}
M.G.~Alford, A.~Schmitt, K.~Rajagopal and T.~Sch{\"a}fer, \emph{{Color superconductivity in dense quark matter}}, \href{https://doi.org/10.1103/RevModPhys.80.1455}{\emph{Rev. Mod. Phys.} {\bfseries 80} (2008) 1455} [\href{https://arxiv.org/abs/0709.4635}{{\ttfamily 0709.4635}}].

\bibitem{Bolognesi:2021hmg}
S.~Bolognesi, K.~Konishi and A.~Luzio, \emph{{Strong anomaly and phases of chiral gauge theories}}, \href{https://doi.org/10.1007/JHEP08(2021)028}{\emph{JHEP} {\bfseries 08} (2021) 028} [\href{https://arxiv.org/abs/2105.03921}{{\ttfamily 2105.03921}}].

\bibitem{Csaki:2021xhi}
C.~Cs\'aki, H.~Murayama and O.~Telem, \emph{{Some exact results in chiral gauge theories}}, \href{https://doi.org/10.1103/PhysRevD.104.065018}{\emph{Phys. Rev. D} {\bfseries 104} (2021) 065018} [\href{https://arxiv.org/abs/2104.10171}{{\ttfamily 2104.10171}}].

\bibitem{Tong:2021phe}
D.~Tong, \emph{{Comments on symmetric mass generation in 2d and 4d}}, \href{https://doi.org/10.1007/JHEP07(2022)001}{\emph{JHEP} {\bfseries 07} (2022) 001} [\href{https://arxiv.org/abs/2104.03997}{{\ttfamily 2104.03997}}].

\bibitem{Razamat:2020kyf}
S.S.~Razamat and D.~Tong, \emph{{Gapped Chiral Fermions}}, \href{https://doi.org/10.1103/PhysRevX.11.011063}{\emph{Phys. Rev. X} {\bfseries 11} (2021) 011063} [\href{https://arxiv.org/abs/2009.05037}{{\ttfamily 2009.05037}}].

\bibitem{Cacciapaglia:2019vce}
G.~Cacciapaglia, S.~Vatani and Z.-W.~Wang, \emph{{Tumbling to the Top}},  \href{https://arxiv.org/abs/1909.08628}{{\ttfamily 1909.08628}}.

\bibitem{Eichten:1981mu}
E.~Eichten and F.~Feinberg, \emph{{Comment on Tumbling Gauge Theories}}, \href{https://doi.org/10.1016/0370-2693(82)91242-4}{\emph{Phys. Lett. B} {\bfseries 110} (1982) 232}.

\bibitem{Wang:2022ucy}
J.~Wang and Y.-Z.~You, \emph{{Symmetric Mass Generation}}, \href{https://doi.org/10.3390/sym14071475}{\emph{Symmetry} {\bfseries 14} (2022) 1475} [\href{https://arxiv.org/abs/2204.14271}{{\ttfamily 2204.14271}}].

\bibitem{Wang:2013yta}
J.~Wang and X.-G.~Wen, \emph{{Nonperturbative regularization of (1+1)-dimensional anomaly-free chiral fermions and bosons: On the equivalence of anomaly matching conditions and boundary gapping rules}}, \href{https://doi.org/10.1103/PhysRevB.107.014311}{\emph{Phys. Rev. B} {\bfseries 107} (2023) 014311} [\href{https://arxiv.org/abs/1307.7480}{{\ttfamily 1307.7480}}].

\bibitem{Wang:2018ugf}
J.~Wang and X.-G.~Wen, \emph{{A Solution to the 1+1D Gauged Chiral Fermion Problem}}, \href{https://doi.org/10.1103/PhysRevD.99.111501}{\emph{Phys. Rev. D} {\bfseries 99} (2018) 111501} [\href{https://arxiv.org/abs/1807.05998}{{\ttfamily 1807.05998}}].

\bibitem{Zeng:2022grc}
M.~Zeng, Z.~Zhu, J.~Wang and Y.-Z.~You, \emph{{Symmetric Mass Generation in the 1+1 Dimensional Chiral Fermion 3-4-5-0 Model}}, \href{https://doi.org/10.1103/PhysRevLett.128.185301}{\emph{Phys. Rev. Lett.} {\bfseries 128} (2022) 185301} [\href{https://arxiv.org/abs/2202.12355}{{\ttfamily 2202.12355}}].

\bibitem{Shtabovenko:2023idz}
V.~Shtabovenko, R.~Mertig and F.~Orellana, \emph{{FeynCalc 10: Do multiloop integrals dream of computer codes?}}, \href{https://doi.org/10.1016/j.cpc.2024.109357}{\emph{Comput. Phys. Commun.} {\bfseries 306} (2025) 109357} [\href{https://arxiv.org/abs/2312.14089}{{\ttfamily 2312.14089}}].

\bibitem{Huber:2019dkb}
M.Q.~Huber, A.K.~Cyrol and J.M.~Pawlowski, \emph{Dofun 3.0: Functional equations in mathematica}, \href{https://doi.org/10.1016/j.cpc.2019.107058}{\emph{Comput. Phys. Commun.} {\bfseries 248} (2020) 107058} [\href{https://arxiv.org/abs/1908.02760}{{\ttfamily 1908.02760}}].

\bibitem{Pawlowski:2021tkk}
J.M.~Pawlowski, C.S.~Schneider and N.~Wink, \emph{{QMeS-Derivation: Mathematica package for the symbolic derivation of functional equations}},  \href{https://arxiv.org/abs/2102.01410}{{\ttfamily 2102.01410}}.

\bibitem{Braun:2025gvq}
J.~Braun, A.~Gei{\ss}el, J.M.~Pawlowski, F.R.~Sattler and N.~Wink, \emph{{Juggling with Tensor Bases in Functional Approaches}},  \href{https://arxiv.org/abs/2503.05580}{{\ttfamily 2503.05580}}.

\bibitem{Gehring:2015vja}
F.~Gehring, H.~Gies and L.~Janssen, \emph{{Fixed-point structure of low-dimensional relativistic fermion field theories: Universality classes and emergent symmetry}}, \href{https://doi.org/10.1103/PhysRevD.92.085046}{\emph{Phys. Rev. D} {\bfseries 92} (2015) 085046} [\href{https://arxiv.org/abs/1506.07570}{{\ttfamily 1506.07570}}].

\bibitem{Braun:2014ata}
J.~Braun, L.~Fister, J.M.~Pawlowski and F.~Rennecke, \emph{From quarks and gluons to hadrons: Chiral symmetry breaking in dynamical qcd}, \href{https://doi.org/10.1103/PhysRevD.94.034016}{\emph{Phys. Rev. D} {\bfseries 94} (2016) 34016} [\href{https://arxiv.org/abs/1412.1045}{{\ttfamily 1412.1045}}].

\bibitem{Jaeckel:2002rm}
J.~Jaeckel and C.~Wetterich, \emph{{Flow equations without mean field ambiguity}}, \href{https://doi.org/10.1103/PhysRevD.68.025020}{\emph{Phys. Rev. D} {\bfseries 68} (2003) 025020} [\href{https://arxiv.org/abs/hep-ph/0207094}{{\ttfamily hep-ph/0207094}}].

\bibitem{Denz:2019ogb}
T.~Denz, M.~Mitter, J.M.~Pawlowski, C.~Wetterich and M.~Yamada, \emph{{Partial bosonization for the two-dimensional Hubbard model}}, \href{https://doi.org/10.1103/PhysRevB.101.155115}{\emph{Phys. Rev. B} {\bfseries 101} (2020) 155115} [\href{https://arxiv.org/abs/1910.08300}{{\ttfamily 1910.08300}}].

\end{thebibliography}\endgroup

\end{document}